\documentclass[aps,prx,twocolumn,superscriptaddress]{revtex4-2}
\usepackage{hyperref}
\usepackage{mathtools}
\hypersetup{
    colorlinks=true,
    linkcolor=blue,
    filecolor=magenta,      
    urlcolor=blue,
    citecolor = red,
    }
\usepackage{lipsum}
\usepackage{amsmath}
\usepackage{amsfonts}
\usepackage{layouts}
\usepackage{graphicx}
\usepackage{lipsum}
\usepackage{comment}
\usepackage{algorithm}
\usepackage{algpseudocode}
\usepackage{tikz}
\usepackage{quantikz}
\usetikzlibrary{quantikz2}
\usepackage{braket}
\usepackage{amssymb}
\usepackage{natbib}
\usepackage{nicematrix}
\usepackage{fancyhdr}

\usepackage{bibunits}

\defaultbibliographystyle{naturemag}   % or nature, unsrt, etc.
\defaultbibliography{refs}

\begin{document}

\title{Reconstructing non-Abelian braiding and fusion without anyon transport}

\author{Lucy Byles}
\affiliation{School of Physics and Astronomy, University of Leeds, Woodhouse Lane, Leeds, LS2 9JT, United Kingdom}

\author{Matthew D. Horner}
\affiliation{Aegiq Ltd., Arundel Street, Sheffield, S1 2NS, United Kingdom}

\author{Benjamin T. H. Varcoe}
\affiliation{School of Physics and Astronomy, University of Leeds, Woodhouse Lane, Leeds, LS2 9JT, United Kingdom}

\author{Jiannis K. Pachos}

\affiliation{School of Physics and Astronomy, University of Leeds, Woodhouse Lane, Leeds, LS2 9JT, United Kingdom}

\date{\today}

\begin{abstract}
Non-Abelian anyons offer a route to fault-tolerant and universal quantum computing, but experimental access to their defining braiding and fusion data remains limited by the resource overhead of implementing extended anyonic processes on quantum hardware. Here we introduce and experimentally realise a measurement-only protocol based on temporally ordered ribbon operations that reconstructs the non-Abelian braiding and fusion primitives of the quantum double model $D(S_3)$ without physical anyon transport. We implement a reduced two-qutrit version of the protocol on Quantinuum's H2 trapped-ion processors, realising ancilla-assisted ribbon operations and anyonic charge projections in a qubit encoding. We reconstruct the squared braiding phases and fusion amplitudes using an adapted Hadamard test and post-selected measurements, respectively. The associated braiding and fusion transformations reproduce their ideal actions with average normalised output-state fidelities of $\overline{\mathcal{F}}_{R}=0.9988$ and $\overline{\mathcal{F}}_{F}=0.9987$. Combining these primitives produces a non-Clifford braid and a non-stabilizer resource state, supporting measurement-only anyonic encodings as building blocks for larger topologically encoded quantum processors.

\end{abstract}

\maketitle

\begin{bibunit}

\section{Topological data without anyon transport}

Topological phases of matter can host anyons, quasiparticles whose exchange statistics go beyond those of bosons and fermions~\cite{Kitaev2003,Nayak2008}. In the non-Abelian case, exchanging and recombining anyons implements non-commuting transformations on a collective Hilbert space, providing a route to quantum information processing in which logical information is stored non-locally and is therefore insensitive to local perturbations~\cite{Freedman2003,Nayak2008}. The defining data of an anyon model are its braiding and fusion transformations, denoted by $R$ and $F$, which determine how worldlines wind around each other and how different fusion bases are related. Experimentally accessing these data is challenging because conventional protocols require controlled creation, motion and coherent fusion measurement of anyonic excitations.

Physical transport, however, is not strictly necessary. In measurement-only topological quantum computation, braiding transformations are generated by sequences of topological charge measurements, using an anyonic analogue of teleportation~\cite{Bonderson2008PRL,Bonderson2009AnnPhys}. This replaces dynamical motion by temporally ordered measurements and projections, a structure well suited to digital quantum hardware where ribbon operators and charge projectors can be implemented directly. Related finite-depth, adaptive and feed-forward constructions have shown that measurement sequences can prepare and manipulate non-Abelian topological order on quantum devices~\cite{BravyiKimKlieschKoenig2022Adaptive,TantivasadakarnVishwanathVerresen2023Hierarchy,VerresenTantivasadakarnVishwanath2021Efficiently,RenTantivasadakarnWilliamson2025SolvableAnyons}.

Kitaev's quantum double models provide an exactly solvable setting in which these ideas can be made explicit~\cite{Kitaev2003}. They belong to a broader class of exactly solvable models for topological order and have a well-developed ribbon-operator and charge-projection formalism~\cite{LevinWen2005StringNet,BombinMartinDelgado2008NonAbelianKitaev,BeigiShorWhalen2011QuantumDoubleBoundary}. For a finite group $G$, the excitations of $D(G)$ are labelled by group-theoretic fluxes and charges, whose manipulations are the basis for topological quantum computation~\cite{OgburnPreskill1999TopologicalQC,Mochon2003NonsolvableAnyons,Mochon2004SmallerGroups}. The smallest group that supports non-Abelian anyons is $S_3$, so $D(S_3)$ is a minimal non-Abelian quantum double with a six-dimensional local Hilbert space. Although $D(S_3)$ braiding alone has finite image and is thus not universal, universality can be recovered when braiding is supplemented by measurements and ancillary resources~\cite{CuiHongWang2015,ChenRen2025}. Equivalently, suitable combinations of braiding and fusion transformations can generate non-Clifford resources, enabling universal computation when combined with Clifford operations and stabilizer measurements~\cite{BravyiKitaev2005,Byles2024}.

Recent experiments on programmable quantum processors and photonic platforms have begun to access non-Abelian anyonic physics in controlled settings~\cite{Andersen2023,Iqbal2024,Xu2024,Goel2024}. Most recently, a large-scale 54-qubit realisation of $D(S_3)$ topological order on a trapped-ion processor encoded logical qutrits in the global fusion space and implemented a universal gate set by combining braiding and fusion operations~\cite{Lo2026}. These advances establish the feasibility of manipulating non-Abelian topological degrees of freedom on quantum hardware. Direct reconstruction of the underlying $R$- and $F$-data, however, remains demanding because the natural $D(S_3)$ lattice description involves local $d=6$ qudits together with non-unitary ribbon and charge-projection operations.

\begin{figure*}[t]
    \centering
    \includegraphics[width=2\columnwidth]{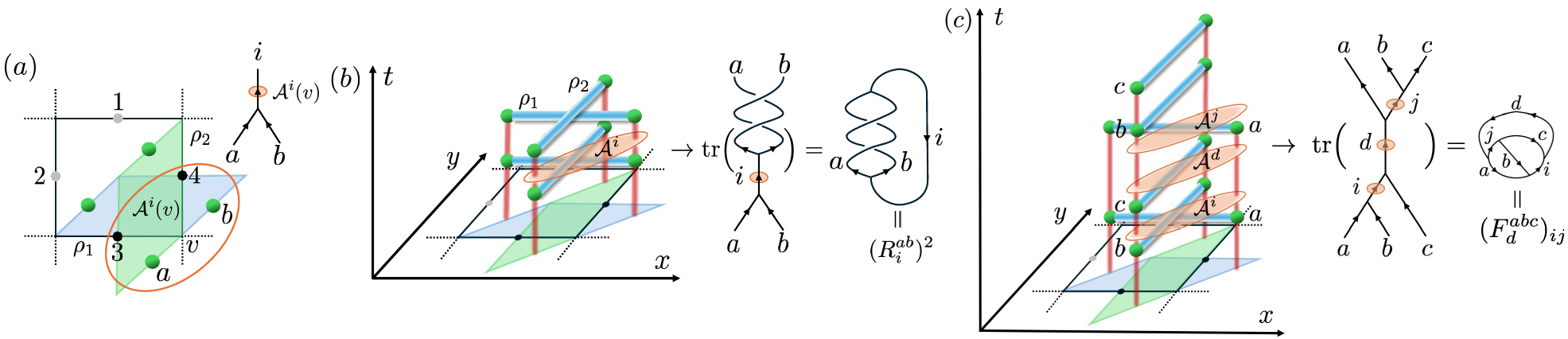}
\caption{{\bf Diagrammatic reconstruction of $(R_i)^2$ and $(F_{ij})^2$.}
(a) Minimal plaquette underlying the reduced two-qutrit protocol. Although the original $D(S_3)$ plaquette contains four $d=6$ qudits, the relevant ribbon and projection operators act non-trivially only on qudits 3 and 4, and their action is subsequently reduced to two qutrits. The reduced ribbons $\mathcal{F}^{G}_{\rho_1}$ and $\mathcal{F}^{G}_{\rho_2}$ create pairs of $G$ anyons (green spheres), and the vertex projector $\mathcal{A}^{i}(v)$ resolves their fusion channel.
(b) Spacetime sequence for reconstructing the squared braiding phase. Temporally ordered ribbon operations and an intermediate charge projection close the worldlines into a double braid of anyons $a$ and $b$ in channel $i$, giving the amplitude $(R^{ab}_{i})^{2}$ without physical anyon transport.
(c) Fusion reconstruction. The ribbon and projection sequence prepares two fusion-basis resolutions of the same anyonic process at one end of the ribbon, giving $(F^{abc}_{d})_{ij}$. Charge neutrality of the ribbons results in the conjugate evolutions at the other end giving overall $(F^{abc}_{d})_{ij}(F^{abc}_{d})^{*}_{ij}=|(F^{abc}_{d})_{ij}|^{2}$, which reduces to $[(F^{abc}_{d})_{ij}]^{2}$ in the real gauge used here.}
    \label{fig:Fig1}
\end{figure*}

Here we address the complementary task of directly reconstructing the underlying $R$- and $F$-data using a compact, observable-specific two-qutrit protocol that provides a modular route towards larger-scale anyon processing. Rather than preparing an extended topologically ordered state and physically transporting anyons, our protocol uses temporally ordered ribbon operations and local charge projections to generate the corresponding braiding and fusion histories. The relevant four-qudit overlaps are then reduced to a two-qutrit implementation, providing a minimal route to characterising the non-Abelian primitives of $D(S_3)$ on qubit-based hardware.

We implement this protocol on Quantinuum's H2 trapped-ion quantum processors~\cite{Moses2023RaceTrackH2}. An adapted Hadamard test reconstructs the common Abelian-sector squared phase $(R^A)^2=(R^B)^2=(R^+)^2$ and the non-Abelian-sector phase $(R^G)^2$, while post-selected overlap and normalisation probabilities determine the squared fusion amplitudes $|F_{ij}|^2$. Combining the reconstructed primitives yields a non-Clifford braid and a non-stabilizer resource state. We further assess the reconstructed transformations through their action on Haar-random states in the logical qutrit fusion space, finding high average output-state fidelities with the corresponding ideal transformations. These results support measurement-only anyonic protocols as building blocks for larger encoded quantum processors.

\section{Reconstructing $D(S_3)$ primitives}

We experimentally reconstruct the braiding and fusion primitives of the minimal non-Abelian quantum double model $D(S_3)$ and show that the reconstructed data are sufficient to generate a non-stabilizer resource state. We focus on the closed fusion subcategory $\{A,B,G\}$, where $A$ is the vacuum, $B$ is an Abelian anyon and $G$ is the non-Abelian anyon. The relevant fusion rules and $R$- and $F$-symbols follow from the $D(S_3)$ quantum-double data and the reduced construction developed in Ref.~\cite{Byles2024} (see also Appendix~\ref{sec:RFqutrit}). The fusion rules
\begin{equation}
    B\times B=A,\quad
    G\times B=G,\quad
    G\times G=A+B+G
\end{equation}
define a three-dimensional fusion space for a pair of $G$ anyons. In the basis
\begin{equation}
    \{
    \ket{G\times G\to A},
    \ket{G\times G\to B},
    \ket{G\times G\to G}
    \},
\label{eqn:fusion}
\end{equation}
the corresponding braiding and fusion transformations are
\begin{equation}
    R =
    \begin{pmatrix}
        \omega & 0 & 0\\
        0 & -\omega & 0\\
        0 & 0 & \bar{\omega}
    \end{pmatrix},
    \quad
    F =
    \frac{1}{2}
    \begin{pmatrix}
        1 & 1 & \sqrt{2}\\
        1 & 1 & -\sqrt{2}\\
        \sqrt{2} & -\sqrt{2} & 0
    \end{pmatrix},
    \label{eq:RF}
\end{equation}
where $\omega=e^{2\pi i/3}$. Experimentally, we access the sectors $i,j\in\{+,G\}$, with $+=A+B$ denoting the composite Abelian sector.

Figure~\ref{fig:Fig1}(a) shows the minimal plaquette geometry underlying the reduced protocol. The full $D(S_3)$ plaquette contains four $d=6$ qudits, but the ribbon and projection operators needed for the present overlaps act non-trivially only on the degrees of freedom labelled 3 and 4. This locality is what renders the dense reduction possible: the relevant four-qudit vacuum expectation values can be evaluated using operators on two reduced qutrits only, greatly simplifying the experimental implementation. The green spheres labelled $a$ and $b$ denote the $G$ anyons created at one end of the two ribbons $\rho_1$ and $\rho_2$. Note that $G$ is self-dual, so the opposite ribbon endpoints also carry $G$ anyons, as required by overall topological charge neutrality. In the quantum-double language these excitations carry both flux and charge data, but for the processes considered here the relevant fusion channel is resolved by the local vertex projector $\mathcal{A}^i(v)$, which distinguishes the $i=+$ and $i=G$ sectors. 

\begin{figure*}
    \centering
\includegraphics[width = \linewidth]{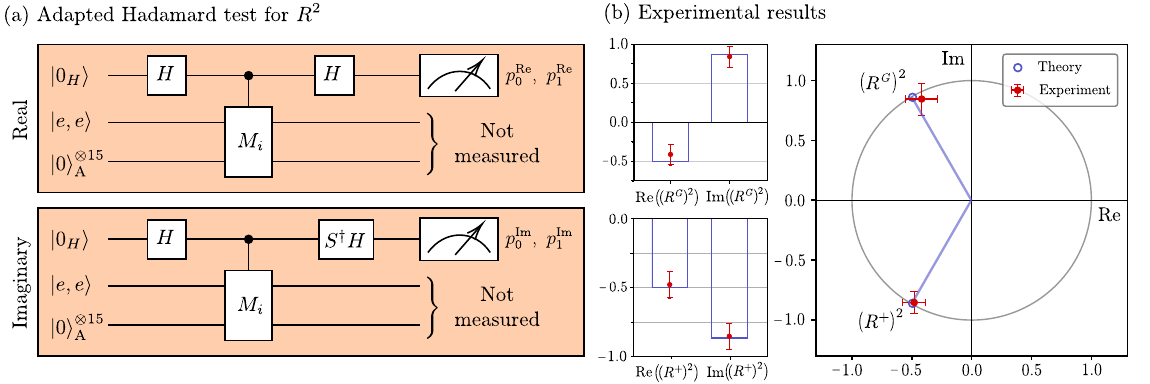}
\caption{{\bf Hardware reconstruction of the squared braiding phases.}
(a) Adapted Hadamard-test circuits used to extract the real and imaginary components of the non-unitary ribbon-projection overlap. The block $M_i$ is the unitary dilation of the reduced two-qutrit map $\mathcal{F}^{G}_{\rho_2}\mathcal{F}^{G}_{\rho_1}\mathcal{A}^{i}(v)\mathcal{F}^{G}_{\rho_2}\mathcal{F}^{G}_{\rho_1}$. The real component is obtained with the final Hadamard measurement basis, while the imaginary component is obtained by inserting $S^\dagger$ before the final Hadamard. The operation ancillas are not measured, as only the Hadamard ancilla outcomes $p_0$ and $p_1$ are used. The $i=+$ circuit gives ${\rm Re}\big((R^+)^2\big)=8(2p^{\rm Re}_0-1)$ and ${\rm Im}\big((R^+)^2\big)=8(2p^{\rm Im}_0-1)$, while replacing $\mathcal{A}^{i}(v)$ by the identity gives $(R^{\mathbf 1})^2$ and hence $(R^G)^2=(R^{\mathbf 1})^2-(R^+)^2$.
(b) Experimental reconstruction of the squared braiding phases. The bar plots show the real and imaginary components of $(R^G)^2$ and $(R^+)^2$, with bars denoting the theoretical values and red markers showing the experimental estimates. The Argand diagram compares the same reconstructed phases with the theoretical roots of unity, $(R^G)^2=\omega$ and $(R^+)^2=\bar{\omega}$. The $+$ sector gives the common squared phase of the $A$ and $B$ fusion channels, $(R^A)^2=(R^B)^2=(R^+)^2$. Error bars denote one standard deviation from binomial shot-noise propagation.}
\label{fig:R2}
\end{figure*}

Successive applications of the ribbon operations and charge projections determine the squared braiding phases and squared fusion amplitudes used below. These elementary building blocks are non-unitary, so their qubit implementation uses ancilla-assisted dilations. In the braiding experiment these dilations are embedded coherently in a Hadamard test, so each shot contributes to the measured quadrature. By contrast, the fusion experiment extracts probabilities from runs that satisfy all required ribbon-ancilla, projection-ancilla and computational-output conditions, which strongly reduces the accepted sample.

The braiding Hadamard-test circuit for the $+$ sector used 22 qubits, with 229 single-qubit and 134 two-qubit gates, and each quadrature was sampled with 7,000 shots. The fusion-overlap circuits used 36 qubits, comprising six computational qubits and 30 ancillas, with 322 single-qubit and 213 two-qubit gates, while the normalisation circuits used 33 qubits, with 275 single-qubit and 184 two-qubit gates. The overlap and normalisation measurements used 11,000 and 13,000 shots, respectively.

\subsection{Braiding phases from ordered ribbons}
\label{sec:R}

Figure~\ref{fig:Fig1}(b) gives the spacetime interpretation of the braiding measurement. The horizontal plane is the plaquette geometry and the vertical direction is time. The ordered application of the ribbons first creates the two $G$ anyons $a$ and $b$, after which the projection $\mathcal{A}^i(v)$ fixes their fusion channel and subsequent ribbon operations are applied. Because the quantity measured is a vacuum expectation value, the initial and final ribbon endpoints are contracted, as indicated by the red vertical lines, producing closed spacetime loops. The ordering of the two final ribbon operations is essential: the sequence shown gives linked loops, corresponding to a double exchange of the anyons $a$ and $b$, whereas the opposite ordering would give unlinked loops and no braid. The closed diagram therefore evaluates to the squared braiding phase $(R^{i})^2$, with $i=+,G$. 

\begin{figure*}[t!]
    \centering
\includegraphics[width = \linewidth]{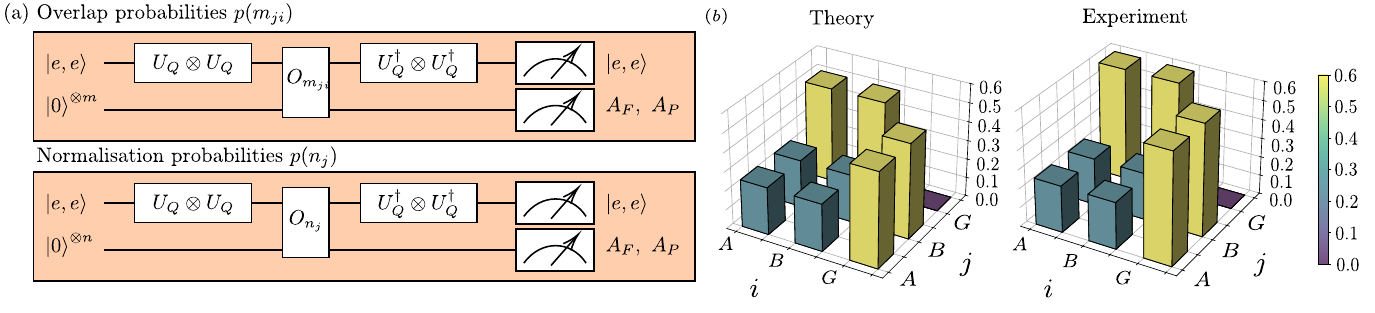}
\caption{{\bf Hardware reconstruction of the squared fusion amplitudes.} 
(a) Post-selected fusion-readout protocol. The upper circuit measures the overlap probabilities $p(m_{ji})$ using $m=30$ ancillas, while the lower circuit measures the normalisation probabilities $p(n_j)$ using $n=27$ ancillas. Unlike the braiding reconstruction, no Hadamard test is used: the probabilities are obtained from bit strings satisfying the ribbon-operation condition $A_F=0\cdots0$, the projection-ancilla syndromes $A_P=(s_i,s_j)$ for the overlap circuit or $s_j$ for the normalisation circuit, and the final computational output $\ket{e,e}$.
(b) Theoretical and experimentally reconstructed matrices of squared fusion amplitudes, $(F_{ij})^2$, in the real gauge used here, with rows and columns ordered as $A,B,G$. The experimental entries are obtained from ratios of the post-selected overlap and normalisation probabilities. The $A$-$B$ symmetry determines the repeated entries. The corresponding matrix $F_{\mathrm{exp}}$ is obtained by taking square roots and fixing the conventional real, symmetric gauge.}
\label{fig:Ftomog}
\end{figure*}

Since $(R^{+})^2$ and $(R^{G})^2$ are complex phases, we reconstruct their real and imaginary components using two variants of an adapted Hadamard test. For the $+$ fusion sector, let $p^{\rm Re}_0$ and $p^{\rm Im}_0$ denote the probabilities of measuring the Hadamard ancilla in $\ket{0}$ in the real- and imaginary-part circuits, respectively. The reconstructed quadratures are then
\begin{equation}
    \operatorname{Re}\!\left[(R^+)^2\right]
    =
    8\left(2p^{\rm Re}_0-1\right),
    \quad
    \operatorname{Im}\!\left[(R^+)^2\right]
    =
    8\left(2p^{\rm Im}_0-1\right).
    \label{eq:Rplus-measurement}
\end{equation}
The real-part circuit uses the standard final Hadamard measurement, whereas the imaginary-part circuit includes an $S^\dagger$ gate before the final Hadamard, as shown in Fig. \ref{fig:R2}(a). The $G$-sector phase is inferred from the identity-channel measurement,
\begin{equation}
    (R^G)^2 = (R^{\mathbf 1})^2-(R^+)^2,
    \label{eq:RG-measurement}
\end{equation}
where $R^{\mathbf 1}$ denotes the same ribbon sequence with the charge projection replaced by the identity. This avoids implementing an additional controlled $G$-projection circuit.

Each Hadamard-test quadrature used $N=7000$ shots. For the $+$ sector we obtain
\begin{equation}
    (R^{+}_{\text{exp}})^{2}
    =
    -0.480\pm0.095
    - i(0.853\pm0.095),
    \label{eqn:R+}
\end{equation}
in agreement with the theoretical value $(R^{+})^{2}=\bar{\omega}\simeq -0.5-i0.866$.
For braiding, this $+$ measurement gives the common squared phase of the $A$ and $B$ fusion channels, since $(R^A)^2=(R^B)^2=(R^+)^2$.
Using Eq.~\eqref{eq:RG-measurement}, we infer
\begin{equation}
    (R^{G}_{\text{exp}})^{2}
    =
    -0.418\pm0.135
    + i(0.846\pm0.134),
    \label{eqn:RG}
\end{equation}
consistent with $(R^{G})^{2}=\omega\simeq -0.5+i0.866$. Figure~\ref{fig:R2}(b) compares the measured and theoretical squared braiding phases in the complex plane. The uncertainties denote one standard deviation from binomial shot-noise propagation, with the uncertainty in $(R^G)^2$ obtained by adding the independent uncertainties in Eq.~\eqref{eq:RG-measurement} in quadrature. In the complex plane, both reconstructed phases are separated by more than five standard deviations from the zero-amplitude value expected from complete depolarisation of the Hadamard ancilla. 

To quantify the performance of the experimentally reconstructed braiding
transformation, $(R_{\mathrm{exp}})^2
    =
    \operatorname{diag}
    \left[
        (R_{\mathrm{exp}}^{+})^2,
        (R_{\mathrm{exp}}^{+})^2,
        (R_{\mathrm{exp}}^{G})^2
    \right]$,
we evaluate its action on states in the logical anyonic fusion space. For a
normalised input state $\ket{\psi}$, the ideal and experimentally
reconstructed output states are
\begin{equation}
    \ket{\psi_{\mathrm{id}}^{R}}
    =
    R^2\ket{\psi},
    \qquad
    \ket{\psi_{\mathrm{exp}}^{R}}
    =
       N_\psi (R_{\mathrm{exp}})^2\ket{\psi},
\end{equation}
with $N_{\psi}=1/ \sqrt{
            \bra{\psi}     [(R_{\mathrm{exp}})^2]^{\dagger}
            (R_{\mathrm{exp}})^2
            \ket{\psi}}$
The output-state fidelity is then
\begin{equation}
    \mathcal{F}_{R}(\psi)
    =
    \left|
        \braket{
            \psi_{\mathrm{id}}^{R}
            |
            \psi_{\mathrm{exp}}^{R}
        }
    \right|^2.
\end{equation}
Averaging $\mathcal{F}_{R}(\psi)$ over $10^5$ Haar-random qutrit input
states gives an average output-state fidelity
\begin{equation}
    \overline{\mathcal{F}}_{R}
    =
    0.9988.
\end{equation}
This quantity assesses the agreement between the normalised ideal and experimental output states. The high fidelity shows that the ancilla-assisted ribbon and projection sequence accurately preserves the relative phases that determine the logical braiding transformation, despite the depth of the circuit. 

\subsection{Fusion amplitudes from post-selection}
\label{sec:F}

Figure~\ref{fig:Fig1}(c) shows the analogous construction for fusion. The lower part of the sequence prepares one fusion-basis resolution of three $G$ anyons: two anyons are first resolved through channel $i$, and the resulting charge subsequently fuses to the final channel $d=G$. Inserting the additional ribbon operation and the projection $\mathcal{A}^j$ changes the order in which the same anyons are fused, producing the second fusion-basis resolution. Taking the vacuum expectation value closes the corresponding worldlines and gives the overlap between these two fusion bases. At one end of the ribbons this overlap is the fusion matrix element $(F^{abc}_{d})_{ij}$. The anyons at the other ribbon endpoints provide the conjugate process, fixed by total charge neutrality, resulting in $(F^{abc}_{d})_{ij}^{*}$. Hence, the experiment reconstructs the closed, charge-neutral diagram that gives the squared fusion amplitudes $|F_{ij}|^2$ from post-selected overlap and normalisation probabilities. 
%\textcolor{purple}{[Note here about getting F from this?]} 
%\textcolor{purple}{Although the experiment reconstructs only the squared elements of the $F$ matrix, we show that this is sufficient to fix its gauge-equivalence class. Taking square roots introduces discrete sign ambiguities, but enforcing reality, unitarity, and symmetry restricts the solutions to four possibilities (up to a global phase). These are all related by gauge transformations corresponding to basis changes in the fusion space. The data therefore uniquely determine $F$ up to gauge, enabling direct comparison with the standard $D(S_{3})$ quantum double model.}
Because the relevant $D(S_3)$ $F$-symbol can be chosen real, the closed fusion diagrams measured here yield $|F_{ij}|^2=(F_{ij})^2$~\cite{MattPeterJJiannisToAppear1}. Taking square roots therefore determines the magnitudes $|F_{ij}|$, while leaving only a discrete ambiguity in their signs. For the symmetric $D(S_3)$ $F$-symbol, compatibility with the unitary fusion transformation restricts these signs to four choices that represent gauge-equivalent forms of the same $F$-symbol. We fix the conventional gauge of Eq.~\eqref{eq:RF} when presenting $F_{\mathrm{exp}}$. Thus, the measured squared elements determine the fusion transformation up to gauge freedom. Owing to finite sampling, the measured magnitudes do not yield an exactly
unitary matrix. 

We denote by $m_{ji}$ the overlap amplitude associated with the branch in which the two charge projections resolve the fusion channels $i$ and $j$, as illustrated in Fig.~\ref{fig:Fig1}(c), and by $n_j$ the corresponding normalisation of a single $j$-resolved branch. The circuits in Fig.~\ref{fig:Ftomog}(a) experimentally determine the associated post-selected probabilities $p(m_{ji})$ and $p(n_j)$, obtained by conditioning on the required outcomes of the ribbon-ancilla $A_F$, projection-ancilla, $A_P$, and computational-output syndromes, $|e,e\rangle$.

The measured post-selected probabilities determine the three independent
squared amplitudes through
\begin{align}
    a &=
    \frac{1}{2}
    \sqrt{\frac{p(m_{++})}{p(n_+)}},
    \nonumber\\
    b &=
    \frac{1}{2}
    \left[
        \sqrt{\frac{p(m_{+G})}{p(n_+)}}
        +
        \frac{1}{2}
        \sqrt{\frac{p(m_{G+})}{p(n_G)}}
    \right],
    \label{eq:abc-extraction}\\
    c &=
    \sqrt{\frac{p(m_{GG})}{p(n_G)}}.
    \nonumber
\end{align}
They give the elementwise squared fusion amplitudes
\begin{equation}
    \big[(F_{ij})^2\big]_{\mathrm{exp}}
    =
    \begin{pmatrix}
        a & a & b\\
        a & a & b\\
        b & b & c
    \end{pmatrix},
\end{equation}
with rows and columns ordered as $A,B,G$.
Figure~\ref{fig:Ftomog}(b) compares these reconstructed quantities with
the corresponding ideal $D(S_3)$ values.

% Taking square roots of these elementwise squared amplitudes and fixing the
% conventional real, symmetric gauge of Eq.~\eqref{eq:RF} gives
% \begin{equation}
%     F_{\mathrm{exp}} =
%     \begin{pmatrix}
%         \sqrt{a} & \sqrt{a} & \sqrt{b}\\
%         \sqrt{a} & \sqrt{a} & -\sqrt{b}\\
%         \sqrt{b} & -\sqrt{b} & \sqrt{c}
%     \end{pmatrix},
%     \label{eq:Phi-form}
% \end{equation}
% where rows and columns are ordered as $A,B,G$. 
% \begin{equation}
%     a=
%     \frac{1}{2}
%     \sqrt{\frac{p(m_{++})}{p(n_+)}},
%     \quad
%     b=
%     \sqrt{\frac{p(m_{+G})}{p(n_+)}},
%     \quad
%     c=
%     \frac{1}{2}
%     \sqrt{\frac{p(m_{G+})}{p(n_G)}} .
%     \label{eq:abc-extraction}
% \end{equation}
Substituting the measured post-selected probabilities into
Eq.~\eqref{eq:abc-extraction}, taking square roots and fixing the
conventional gauge gives
\begin{equation}
    F_{\mathrm{exp}}
    =
    \begin{pmatrix}
        0.497 \pm 0.022 & 0.497 \pm 0.022 & 0.768 \pm 0.062\\
        0.497 \pm 0.022 & 0.497 \pm 0.022 & -0.768 \pm 0.062\\
        0.768 \pm 0.062 & -0.768 \pm 0.062 & 0
    \end{pmatrix}
    % \begin{pmatrix}
    %     0.247 \pm 0.0223 & 0.247 \pm 0.0223 & 0.591 \pm 0.0749\\
    %     0.247 \pm 0.0223 & 0.247 \pm 0.0223 & 0.591 \pm 0.0749\\
    %     0.591 \pm 0.0749 & 0.591 \pm 0.0749 & 0.000 \pm 0.0000
    % \end{pmatrix},
    \label{eq:Phi-exp}
\end{equation}
% $\Phi_{AA}=\Phi_{AB}=\Phi_{BA}=\Phi_{BB}=1/4$, $\Phi_{AG}=\Phi_{BG}=\Phi_{GA}=\Phi_{GB}=1/2$ and $\Phi_{GG}=0$
to be compared with the ideal values $F_{AA}=F_{AB}=F_{BA}=F_{BB}=1/2$, $F_{AG}=F_{GA}=1/\sqrt{2}$, $F_{BG}=F_{GB}=-1/\sqrt{2}$ and $F_{GG}=0$. The quoted uncertainties denote one standard deviation obtained by propagating binomial sampling noise through Eq.~\eqref{eq:abc-extraction}. The repeated magnitudes in the $A$ and $B$ rows follow from the
$A$-$B$ symmetry, while their relative signs are fixed by the conventional gauge. 
The dominant uncertainty occurs in the off-diagonal entries involving the
$G$ channel, owing to the small normalisation probability
$p(n_G)=4.6\times10^{-4}$. This post-selection retains only $0.046\%$ of the corresponding trials: for $10^4$ shots one expects about $5$ accepted events. The resulting order-unity relative fluctuation in this channel is amplified through the square-root ratio in Eq.~\eqref{eq:abc-extraction}, giving the larger uncertainty in $b=(F_{AG})^{2}=(F_{GA})^{2}=(F_{BG})^{2}=(F_{GB})^{2}$. The absence of accepted $p(m_{GG})$ events is consistent with the theoretical condition $F_{GG}=0$ as seen from the exact
$D(S_3)$ $F$-symbol given in Eq.~\eqref{eq:RF}. Finally, the other admissible sign assignments correspond to gauge-equivalent representations and do not change the physical conclusions.

%As with the braiding matrix, we quantify the accuracy of this experimentally obtained matrix $F_{\mathrm{exp}}$ through its action on states within the logical anyonic fusion space. Taking a Haar randomised sample of $10^5$ such states, the operational fidelity is calculated to be $0.9987$ to 4 significant figures.

As for the braiding transformation, we quantify the accuracy of the
experimentally reconstructed fusion matrix $F_{\mathrm{exp}}$ through its
action on the logical anyonic fusion space. For each normalised Haar-random
qutrit input state $\ket{\psi}$, we compare the ideal output
$F\ket{\psi}$ with the normalised reconstructed output
\begin{equation}
    \ket{\psi_{\mathrm{exp}}^{F}} = \frac{F_{\mathrm{exp}}\ket{\psi}} {\sqrt{\bra{\psi}F_{\mathrm{exp}}^{\dagger}
    F_{\mathrm{exp}}\ket{\psi}}}.
\end{equation}
Averaging their squared overlap over $10^{5}$ independently sampled input
states gives an average output-state fidelity
\begin{equation}
    \overline{\mathcal{F}}_{F}=0.9987.
\end{equation}
This high conditional fidelity is consistent with the heralding and
code-space filtering built into the fusion protocol. Outcomes are retained
only when the ribbon ancillas, charge-sector flags and final computational
register all satisfy the prescribed conditions. Moreover, the
unit-Hamming-weight encoding maps a single physical bit flip outside the
logical qutrit subspace, allowing many faulty trajectories to be rejected
rather than incorporated into the reconstructed logical data. These
filtering mechanisms preserve the accuracy of the accepted reconstruction
despite the long sequence of ribbon and charge-projection operations, at the
cost of a low acceptance probability and increased statistical uncertainty.

\subsection{A non-Clifford anyonic resource}

Having reconstructed the squared braiding phases and fusion amplitudes, we
combine these primitives into the composite braid
\begin{equation}
    B_s = F_s^{\dagger}R^2F_s,
    \label{eq:Braid-composite}
\end{equation}
where $F_s$ denotes one of the fusion matrices compatible with the measured
squared amplitudes. Since the fusion experiment determines only
$|F_{ij}|^2$, a discrete sign ambiguity remains in reconstructing
$F$. We therefore impose the symmetry and unitarity of the $D(S_3)$ $F$-symbol, 
% together with the condition $F_{GG}=0$ (equivalently, $d=\Phi_{GG}=0$), 
reducing the allowed reconstructions to four admissible sign assignments up to a global phase. However, when the corresponding braids act on the vacuum fusion state $\ket A$, these four admissible sign assignments yield only two distinct output states, differing solely by the relative sign of the $\ket B$ component. Since these two states have the same positive stabilizer R\'enyi entropy, the resulting braid is non-Clifford for every allowed sign choice.

Applying the reconstructed braid up to gauge freedom to the vacuum fusion state gives
\begin{equation}
    \begin{split}
    \ket{\psi^{B}_{\mathrm{exp}}}
    \propto \big[2a(R^{+})^2+b(R^{G})^2\big]\ket A \qquad\\
    \quad \pm 
    \big[2a(R^{+})^2-b(R^{G})^2\big]\ket B,
    \end{split}
    \label{eq:magic-state}
\end{equation}
where here $R^+$ and $R^G$ denote the corresponding braiding phases entering the composite braid, and the sign reflects the residual freedom in reconstructing $F$. We quantify the non-stabilizer character of this state using the second-order stabilizer R\'enyi entropy $M_2$, a magic monotone for pure states~\cite{LeoneOlivieroHamma2022StabilizerRenyi,HaugPiroli2023StabilizerEntropies,LeoneBittel2024StabilizerMonotones}. The ideal state has
\begin{equation}
    M_2(\ket{\psi^{B}_{\mathrm{th}}})
    =
    \log\!\left(\frac{16}{13}\right)
    \simeq
    0.208 .
\end{equation}
From the experimentally reconstructed parameters we obtain
\begin{equation}
    M_2(\ket{\psi^{B}_{\mathrm{exp}}})
    =
    0.258\pm0.0886 .
    \label{eq:M2-exp}
\end{equation}
The uncertainty is estimated by Monte Carlo propagation over the measured inputs, which accounts for the correlations introduced by normalising the reconstructed state. The two relative-sign choices in Eq.~\eqref{eq:magic-state} have identical stabilizer R\'enyi entropy, so all valid sign-gauge choices give the same value of $M_2$. Since $\ket A$ is a stabilizer state and Clifford operations preserve the set of stabilizer states, the generation of a state with $M_2>0$, demonstrates non-stabilizer resource generation and supports the non-Clifford character of the corresponding ideal braid.

%directly certifies that every braid $B_s$ compatible with the measured fusion amplitudes is non-Clifford. 
%The non-Clifford conclusion is therefore independent of the unresolved signs of the reconstructed $F$ matrix.

Together, these results establish a compact hardware building block for programmable non-Abelian topological processing. The experimentally reconstructed braiding and fusion transformations achieve average normalised output-state fidelities of $\overline{\mathcal F}_{R}=0.9988$ and $\overline{\mathcal F}_{F}=0.9987$, respectively, while their combination generates a non-stabilizer resource state. In contrast to approaches that prepare an extended $D(S_3)$ topological state and encode information globally in its fusion space~\cite{Lo2026}, our construction compresses the relevant plaquette histories to a two-qutrit logical protocol. This reduced representation lowers the state-preparation overhead and allows the elementary braiding and fusion primitives to be characterised with high accuracy before larger structures are assembled.

The present experiment addresses the topological data of a single encoded plaquette. A natural next step is to couple several such modules and apply ribbon and charge-projection operations between them, thereby enlarging the fusion space and enabling genuine multi-anyon processes. The combination of a compact logical representation, high-fidelity elementary primitives and a modular measurement-only architecture provides a promising route towards scaling braiding, fusion, magic-state preparation and logical gates within a larger encoded anyonic processor.

\section{Methods}

The Methods are organised as follows. We first present the dense reduction that makes the experimental implementation feasible. Starting from the minimal four-qudit plaquette of the $D(S_3)$ quantum double, we show that the ribbon and charge-projection overlaps needed to reconstruct the braiding and fusion data can be reduced to expectation values of operators acting on only two qutrits. We further show that these expectation values can be evaluated on a simple product state, rather than on the full plaquette ground state, thereby producing a protocol suited to directly encode on the available quantum hardware.

We then describe the qubit implementation of this reduced two-qutrit protocol. Each qutrit is encoded in the unit-Hamming-weight subspace of three physical qubits, and the required qutrit unitaries are decomposed into one- and two-qubit gates. The reduced ribbon and projection operators are non-unitary, so their implementation requires ancilla-assisted heralded operations. We construct these primitives with an economical use of ancillas and specify how successful operation is identified by measurement outcomes. Finally, we explain how the resulting circuits are combined on trapped-ion hardware: an adapted Hadamard test is used to extract the real and imaginary parts of the squared braiding phases, while the squared fusion amplitudes are reconstructed from post-selected overlap and normalisation probabilities.

\subsection{Dense reduction from four qudits to two qutrits}
\label{sec:dense}

We first describe the reduction that makes the experimental implementation feasible. The braiding and fusion data of the $D(S_3)$ model can be extracted from overlaps of ribbon and charge-projection operators acting on the ground state $\ket{\eta}$ of a minimal four-qudit plaquette, as shown in Fig.~\ref{fig:Fig1}(a). Working within the reduced sector $\{+,G\}$, for the braiding phases one has
\begin{equation}
    (R ^{i})^{2} = N_{i}\bra{\eta}F^{G}_{\rho_{2}}F^{G}_{\rho_{1}}A^{i}(v)F^{G}_{\rho_{2}}F^{G}_{\rho_{1}}\ket{\eta},
    \quad i=+,G,
    \label{eq:R2}
\end{equation}
where $(R^{A})^{2}=(R^{B})^{2}=(R^{+})^{2}$ and $A^{+}(v)=A^{A}(v)+A^{B}(v)$ defined as the combined charge projection operator. For both sectors, the four-qudit normalisation factor is $N_{+}=N_{G}=2$. Here $+$ denotes the combined charge sector $A+B$. This should be distinguished from the normalised superposition $\ket{+}=(\ket A+\ket B)/\sqrt{2}$ used when describing states in the fusion space.

Similarly, the fusion amplitudes may be reconstructed from two sets of measured expectation values
\begin{equation}
    \begin{split}
        \bra{\eta}O_{m_{ji}}\ket{\eta}= \qquad\qquad\hspace{3.3cm} \\
        \bra{\eta}F^{G}_{\rho_{2}}F^{G}_{\rho_{2}}A^{j}(v)F^{G}_{\rho_{1}}A^{G}(v)F^{G}_{\rho_{2}}A^{i}(v)F^{G}_{\rho_{2}}F^{G}_{\rho_{1}}\ket{\eta},
    \end{split}
    \label{eq:F21}
\end{equation}
and
\begin{equation}
    \begin{split}
        \bra{\eta}O_{n_{j}}\ket{\eta}= \qquad\qquad\hspace{3.5cm} \\
        \bra{\eta}F^{G}_{\rho_{2}}F^{G}_{\rho_{2}}A^{j}(v)F^{G}_{\rho_{1}}A^{G}(v)F^{G}_{\rho_{1}}A^{j}(v)F^{G}_{\rho_{2}}F^{G}_{\rho_{2}}\ket{\eta},
    \end{split}
    \label{eq:F22}
\end{equation}
% in Appendix \ref{sec:Fqutrit} it is shown 
% Similarly, the fusion amplitudes
% while the fusion amplitudes are obtained from the corresponding overlap between two fusion bases,
% \begin{equation}
%     \begin{split}
%         (F_{ij})^2 = \qquad\qquad\hspace{4cm}\\
%         M_{ij}\left|\frac{\bra{\eta}F^{G}_{\rho_{2}}F^{G}_{\rho_{2}}A^{j}(v)F^{G}_{\rho_{1}}A^{G}(v)F^{G}_{\rho_{2}}A^{i}(v)F^{G}_{\rho_{2}}F^{G}_{\rho_{1}}\ket{\eta}}{\bra{\eta}F^{G}_{\rho_{2}}F^{G}_{\rho_{2}}A^{j}(v)F^{G}_{\rho_{1}}A^{G}(v)F^{G}_{\rho_{1}}A^{j}(v)F^{G}_{\rho_{2}}F^{G}_{\rho_{2}}\ket{\eta}}\right|,
%     \end{split}
%     \label{eq:F2}
% \end{equation}
with $i,j\in\{+,G\}$ as outlined in Appendix \ref{sec:Fqutrit}.
% Here $N_i$ and $M_{ij}$ are normalisation constants, and $i,j\in\{+,G\}$ with $+=A+B$. 
When the fusion amplitude is reconstructed, the modulus of the normalised overlaps is taken, removing the known intermediate braiding phase generated by the intersecting ribbons. The resulting quantity is the squared fusion amplitude $|F_{ij}|^2$.
%We note that for  Eq.~\eqref{eq:F21} the absolute value is taken to remove the intermediate braiding phase generated by overlapping ribbon operators. Subsequently, the measured quantity is the squared fusion amplitude.

The operators comprising Eqs.~\eqref{eq:R2}$-$\eqref{eq:F22} act non-trivially on only two of the four plaquette qudits~\cite{Byles2024}. Denoting a general such operator by $O=\mathbf{1}_{36}\otimes o$, the corresponding expectation value with respect to the four-qudit ground state reduces to a sum of two-qudit expectation values,
\begin{equation}
\bra{\eta} O \ket{\eta} = \bra{\eta} (\mathbf{1}_{36} \otimes o) \ket{\eta} = 6 \sum_{g \in S_3} \bra{\psi_g} o \ket{\psi_g},
\label{eq:reduction-to-2qudits}
\end{equation}
where
\begin{equation}
\ket{\psi_g} = \frac{1}{\sqrt{216}} \sum_{g_1 g_2 = g} \ket{g_1, g_2},
\label{eq:psig-state}
\end{equation}
where $g_1,g_2 \in S_3$. With this convention the states $\ket{\psi_g}$ are not individually normalised. Their normalisation is included in the prefactor in Eq.~\eqref{eq:reduction-to-2qudits}. This establishes the first reduction, from the four-qudit plaquette to two $d=6$ qudits.

The second reduction uses the group structure $S_3\simeq \mathbb{Z}_3\rtimes \mathbb{Z}_2$~\cite{dummit2004abstract}. In a basis ordered as $\{e,c,c^2\}\oplus\{t,tc,tc^2\}$, the relevant operators in the $\{+,G\}$ sector have support only on the $\mathbb{Z}_3$ block, while the complementary block cancels in the group-averaged overlaps. The four-qudit matrix elements can therefore be computed from reduced two-qutrit operators acting on the subspace $\mathbb{Z}_3=\{e,c,c^2\}$. These operators are
\begin{align}
    & \mathcal{F}^{G}_{\rho_{1}} = Z\otimes X + Z^{2} \otimes X^{2}, \label{eq:fg13}\\
    & \mathcal{F}^{G}_{\rho_{2}} = X\otimes Z^{2} + X^{2}\otimes Z, \label{eq:fg23}\\
    & \mathcal{A}^{+}(v) = \frac{1}{3} [\mathbf{1}_{9} + X\otimes X^{2} + X^{2}\otimes X], \label{eq:ap3}\\ 
    & \mathcal{A}^{G}(v) = \frac{1}{3} [2\cdot \mathbf{1}_{9} - X\otimes X^{2} - X^{2}\otimes X ], \label{eq:ag3}
\end{align}
where
\begin{equation}
    X = \begin{pmatrix}
        0 & 0 & 1 \\
        1 & 0 & 0 \\
        0 & 1 & 0
    \end{pmatrix}, \hspace{0.5cm} Z = \begin{pmatrix}
        1 & 0 & 0 \\
        0 & \omega & 0 \\
        0 & 0 & \bar{\omega}
    \end{pmatrix}.
\end{equation}

The operators $\mathcal{F}^{G}_{\rho_1}$ and $\mathcal{F}^{G}_{\rho_2}$ are the reduced ribbon operators, while $\mathcal{A}^{+}(v)$ and $\mathcal{A}^{G}(v)$ project onto the $+$ and $G$ charge sectors. Their fusion and braiding properties are verified in Appendix~\ref{sec:RFqutrit}. 

Finally, the required two-qutrit expectation values can be evaluated on a product state rather than on the entangled states $\ket{\Psi_g}$. Define
\begin{equation}
\ket{\xi} = \frac{1}{3} \sum_{g_1, g_2 \in \mathbb{Z}_3} \ket{g_1, g_2}.
\label{eq:xi}
\end{equation}
For the operator products appearing in Eqs.~\eqref{eq:R2}$-$ \eqref{eq:F22}, the flux-nonconserving matrix elements may be shown to cancel in the sum over $\mathbb{Z}_3$ (see Appendix \ref{sec:densePROD}). The original four-qudit overlaps are therefore proportional to expectation values of the reduced operators on $\ket{\xi}$ as
\begin{equation} \bra{\eta}O\ket{\eta} = \frac{1}{4} \bra{\xi}\mathcal{O}\ket{\xi}, \end{equation} 
where $\mathcal{O}$ denotes the corresponding product of reduced two-qutrit operators. This sequence enables us to reconstruct the relevant squared braiding phases
and fusion amplitudes while replacing the four-qudit plaquette by two qutrits
prepared in the simple product state $\ket{\xi}$.

% The braiding phases in this dense encoding scheme are thus derived as
% \begin{equation}
%     (R ^{i})^{2} = \mathcal{N}_{i}\bra{\xi}\mathcal{F}^{G}_{\rho_{2}}\mathcal{F}^{G}_{\rho_{1}}\mathcal{A}^{i}(v)\mathcal{F}^{G}_{\rho_{2}}\mathcal{F}^{G}_{\rho_{1}}\ket{\xi},
%     \quad i=+,G,
%     \label{eq:R2q}
% \end{equation}
% with normalization $\mathcal{N}_{+}=\mathcal{N}_{G}=1/2$.

% The reduced $+$ and $G$ measurements are sufficient to recover the corresponding $A$, $B$ and $G$ data. For braiding, $(R^A)^2=(R^B)^2=(R^+)^2$, while $(R^G)^2$ is obtained from the identity-channel subtraction. For fusion, the accessible $+$ and $G$ overlaps, together with the $A$-$B$ symmetry and unitarity constraints, determine the squared matrix elements $\Phi_{ij}=|F_{ij}|^2$ in the $A,B,G$ basis (see Appendix \ref{sec:Fqutrit}).

\subsection{Qubit implementation of the reduced two-qutrit protocol}
\label{sec:circuits}

The reduction described above expresses the braiding and fusion data as expectation values of products of two-qutrit ribbon and charge-projection operators. We now describe how these reduced operators are implemented on Quantinuum's trapped-ion quantum processor. Since the hardware operates on qubits, each logical qutrit is encoded into the unit-Hamming-weight subspace of three physical qubits. The two-qutrit protocol is therefore realised on six computational qubits, supplemented by ancilla qubits used to implement non-unitary primitives.

The operators required for the protocol fall into two classes. The qutrit Fourier transforms and basis changes are unitary and can be compiled into two-qubit Givens rotations acting within the encoded qutrit subspace. By contrast, the reduced charge projectors and ribbon operators are non-unitary. These are implemented probabilistically by coupling the computational qubits to ancillas and post-selecting on specified measurement outcomes. The resulting heralded operations realise the desired non-unitary maps on the logical qutrit subspace, with success probabilities that depend on the input state.

The following subsections describe the construction in stages. We first introduce the unit-Hamming-weight qutrit encoding and the Givens rotations used to compile single-qutrit unitaries. We then give the circuit for preparing the product input state $\ket{\xi}$, followed by the ancilla-assisted implementations of the projection and ribbon operators. Finally, we explain how these circuits are combined experimentally: the real and imaginary components of the squared braiding phases are extracted using an adapted Hadamard test, while the fusion amplitudes are reconstructed from post-selected overlap probabilities and their corresponding normalisation factors.

\subsubsection{Qutrit encoding and Givens rotations}

The reduced protocol of Sec.~\ref{sec:dense} is naturally formulated in terms of two qutrits, whereas the Quantinuum processor operates on qubits. We therefore encode each qutrit into the unit-Hamming-weight subspace of three qubits,
\begin{equation}
\begin{aligned}
    \ket{e} \equiv \ket{1,0,0},\,\,
    \ket{c} \equiv \ket{0,1,0},\,\,
    \ket{c^2} \equiv \ket{0,0,1}.
\end{aligned}
\label{eq:encode}
\end{equation}
This three-dimensional subspace defines the logical qutrit Hilbert space. We label logical qutrit states by group elements $g\in \mathbb{Z}_3$, while physical qubit states are labelled by binary occupations $n\in\{0,1\}$. A useful feature of this encoding is that any single-qubit bit-flip error maps a valid code state outside the unit-Hamming-weight subspace and is therefore detectable by a code-space check. Logical bit-flip errors require at least two physical bit flips, so the encoding has effective distance $d=2$ against this error type.

To implement a $\mathrm{U}(3)$ single-qutrit unitary, we decompose it into a product of $\mathrm{U}(2)$ unitary rotations, known as Givens rotations, that act non-trivially on two-dimensional subspaces of the three-dimensional qutrit Hilbert space. In the qubit encoding, these subspaces correspond to pairs of qubits and each Givens rotation is realised as a two-qubit gate acting on the unit-Hamming-weight subspace spanned by $\ket{01}$ and $\ket{10}$, while leaving $\ket{00}$ and $\ket{11}$ unchanged. 
% Equivalently, in the qubit encoding these are two-qubit gates acting on the unit-Hamming-weight subspace spanned by $\ket{01}$ and $\ket{10}$, while leaving the complementary states $\ket{00}$ and $\ket{11}$ unchanged. 
A general such gate has the block form
\begin{equation}
    U =
    \begin{pmatrix}
        1 & 0 & 0 & 0\\
        0 & \alpha & \beta & 0\\
        0 & \gamma & \delta & 0\\
        0 & 0 & 0 & 1
    \end{pmatrix},
    \label{eq:U2}
\end{equation}
where $|\alpha|^2+|\beta|^2=|\gamma|^2+|\delta|^2=1$ and $\alpha\gamma^*+\beta\delta^*=0$. These gates realise the $\mathrm{U}(2)$ rotations appearing in standard decompositions of $\mathrm{U}(3)$ operations, such as the Reck or Clements decompositions~\cite{reck1994experimental, Clements:16}. They therefore provide the elementary building blocks for compiling the qutrit Fourier transform and the basis changes used in the ribbon and projection circuits as we will see in the following.

% \textcolor{gray}{The operators needed for the protocol are not all unitary. In particular, the reduced charge projectors and ribbon operators are implemented as non-unitary primitives by coupling to ancilla qubits and post-selecting on specified measurement outcomes. We first describe the unitary preparation of the two-qutrit input state and then turn to the ancilla-assisted implementations of the projection and ribbon operators.}

\subsubsection{Generalised Fourier transform and initial state preparation}

The state $\ket{\xi}$ defined in Eq.~\eqref{eq:xi} is the equal superposition over all two-qutrit basis states. Since it factorises into a product of two identical single-qutrit superpositions, it can be prepared by applying the qutrit Fourier transform to each qutrit as
\begin{equation}
    \ket{\xi}
    =
    \left(U_{\mathrm{Q}}\otimes U_{\mathrm{Q}}\right)
    \ket{e,e}
    =
    \frac{1}{3}
    \sum_{g_1,g_2\in \mathbb{Z}_3}
    \ket{g_1,g_2},
    \label{eq:xi-preparation}
\end{equation}
where $U_\text{Q}$ is the single-qutrit Fourier transform. In the basis $\{\ket{e},\ket{c},\ket{c^2}\}$ this is given by
\begin{equation}
   U_{\mathrm{Q}}
   =
   \frac{1}{\sqrt{3}}
   \begin{pmatrix}
       1 & 1 & 1\\
       1 & \omega & \omega^2\\
       1 & \omega^2 & \omega
   \end{pmatrix},
   \label{eq:qutrit_qft}
\end{equation}
where $\omega=e^{2\pi i/3}$.

To compile $U_{\mathrm{Q}}$ into the three-qubit encoding of Eq.~\eqref{eq:encode}, we decompose it into Givens rotations using Reck's decomposition~\cite{reck1994experimental}. Each Givens rotation is then implemented as a two-qubit gate acting on the unit-Hamming-weight subspace, as described in Eq.~\eqref{eq:U2}. Together with single-qubit $z$-rotations, these gates implement the qutrit Fourier transform within the logical qutrit subspace while acting trivially on the states outside the relevant two-dimensional rotation subspaces. In Appendix~\ref{app:QFT} we show the qubit circuits for this gate.

\subsubsection{Projection operators}
\label{sec:proj}

The reduced charge operators $\mathcal{A}^{+}(v)$ and $\mathcal{A}^{G}(v)$ are non-unitary and are therefore implemented using ancilla-assisted post-selection. It is convenient to first diagonalise them using the qutrit Fourier transform. Let $P^+$ denote the projector onto the two-qutrit subspace with equal qutrit labels, $g_1=g_2$, given by
\begin{equation}
    P^+\ket{g_1,g_2}
    =
    \begin{cases}
        \ket{g_1,g_2}, & g_1=g_2\\
        0, & g_1\neq g_2
    \end{cases}.
\end{equation}
Moreover, let $P^G=\mathbf{1}_9-P^+$ denote the complementary projection. The reduced charge operators are obtained by conjugating these diagonal projectors with Fourier transforms as
\begin{align}
    \mathcal{A}^{+}(v)
    & =
    (U_{\mathrm{Q}}\otimes U_{\mathrm{Q}})
    P^+
    (U_{\mathrm{Q}}^{\dagger}\otimes U_{\mathrm{Q}}^{\dagger}),
    \label{eq:Aplus-diagonal} \\
    \mathcal{A}^{G}(v)
    & =
    (U_{\mathrm{Q}}\otimes U_{\mathrm{Q}})
    P^G
    (U_{\mathrm{Q}}^{\dagger}\otimes U_{\mathrm{Q}}^{\dagger}).
    \label{eq:AG-diagonal}
\end{align}
The projectors $P^i$ can be implemented with the aid of ancillas, measurement and post-selection. In Appendix~\ref{app:projection_operators} we present the qubit circuits that implement these projectors.

\subsubsection{Ribbon operators}

The reduced ribbon operators $\mathcal{F}^{G}_{\rho_1}$ and $\mathcal{F}^{G}_{\rho_2}$ are non-unitary two-qutrit operators. As with the charge projections, it is useful to diagonalise them by single-qutrit Fourier transforms as 
\begin{align}
    \mathcal{F}^{G}_{\rho_1}
    &=
    (\mathbf{1}_3\otimes U_{\mathrm{Q}})
    \mathcal{D}_{F}
    (\mathbf{1}_3\otimes U_{\mathrm{Q}}^{\dagger}),
    \label{eq:ribbon-diagonal-1}\\
    \mathcal{F}^{G}_{\rho_2}
    &=
    (U_{\mathrm{Q}}^{\dagger}\otimes \mathbf{1}_3)
    \mathcal{D}_{F}
    (U_{\mathrm{Q}}\otimes \mathbf{1}_3),
    \label{eq:ribbon-diagonal-2}
\end{align}
where $U_{\mathrm{Q}}$ is defined in Eq.~\eqref{eq:qutrit_qft}. The diagonal operator has eigenvalues
\begin{equation}
    \frac{1}{2}\mathcal{D}_{F}\ket{g_1,g_2}
    =
    \begin{cases}
        \ket{g_1,g_2}, & g_1=g_2,\\
        -\frac{1}{2}\ket{g_1,g_2}, & g_1\neq g_2,
    \end{cases}
    \label{eq:DF-unscaled}
\end{equation}
where the overall factor of $\frac{1}{2}$ is included for ease of direct circuit implementation. This diagonal gate can be implemented with the aid of ancillas, measurement and post-selection. In Appendix~\ref{app:ribbon_operators} we present the qubit circuits that implements this gate.

\subsection{Experimental implementation on trapped-ion hardware}

We implemented the reduced two-qutrit protocol on Quantinuum's H2 trapped-ion processors. Each logical qutrit is encoded in the unit-Hamming-weight subspace of three physical qubits, so that the two-qutrit computational register uses six qubits. The remaining qubits are ancillas used either to realise the non-unitary ribbon and projection primitives through unitary dilations, or to record the syndromes that identify successful operations and computational outputs. This architecture leads to two distinct experimental readout modes. For braiding, the non-unitary primitives are embedded coherently inside an adapted Hadamard test, so the operation ancillas are not post-selected during the experiment and every shot contributes to the measured quadrature. For fusion, by contrast, the desired quantities are reconstructed from post-selected overlap and normalisation probabilities, requiring simultaneous conditions on ribbon ancillas, projection ancillas and the final computational state. The fusion reconstruction therefore uses larger circuits and a much smaller effective sample size, which is the dominant source of the statistical uncertainty reported in the main text.

% \textcolor{gray}{To implement this code experimentally, we used the Quantinuum H2-2 quantum processor which allows for circuits up to 56 qubits.}

% \textcolor{gray}{Implementing eq \ref{eq:F2} used 36 qubits and 571 total gates, 322 single qubit gates and 213 two qubit gates.
% The final probability distribution was obtained by combining the results of 11000 experimental runs. 
% Implementing eq \ref{other F2} used 33 qubits and 492 total gates, comprising 275 single qubit gates and 184 two qubit gates. }

%\textcolor{gray}{Implementation Challenges}

%\textcolor{gray}{Assessing the outcome}

\subsubsection{Experimental extraction of the braiding phases}
\label{sec:braiding-extraction}

Having constructed circuits that realise the two-qutrit ribbon and projection operators, we now outline the method by which these operations may be used to reconstruct the braiding phases on Quantinuum's H2 quantum processors.
%we now show how these operations may be used to reconstruct the $R$ and $F$ matrices --

Following the dense encoding of Sec.~\ref{sec:dense} the squared braiding elements $(R^{i})^{2}$ are reconstructed as the inner product of a controlled combination of ribbon and projection operators on the reduced input state $\ket{\xi}$ as
\begin{equation}
    (R ^{i})^{2} = \mathcal{N}_{i}\bra{\xi}\mathcal{F}^{G}_{\rho_{2}}\mathcal{F}^{G}_{\rho_{1}}\mathcal{A}^{i}(v)\mathcal{F}^{G}_{\rho_{2}}\mathcal{F}^{G}_{\rho_{1}}\ket{\xi},
    \quad i=+,G,
    \label{eq:R2q}
\end{equation}
with normalization $\mathcal{N}_{+}=\mathcal{N}_{G}=1/2$.
 As these braid group elements are in general complex phase factors, we require a method by which both the real and imaginary components may be extracted from these expectation values. 
%To this end we employ an adapted version of the Hadamard test [ref].
%The outline for this protocol is shown on the left of Fig.().

% For all non-unitary operations $\mathcal{A}^{+}(v), \mathcal{F}^{G}_{\rho_{1}}$ and $\mathcal{F}^{G}_{\rho_{2}}$ we have outlined an encoding whereby the non-unitary action of the respective operation is embedded within a higher-dimensional unitary circuit and implemented probabilistically utilising post-selection on a number of ancillas. Generalising, we thus denote any two-qutrit product of such operations as the non-unitary map $\mathcal{M}:\ket{\psi}\mapsto\ket{\phi}$
% We begin by denoting some general non-unitary two-qutrit operator $\mathcal{M}$ as the map $\mathcal{M}:\ket{\psi}\mapsto\ket{\phi}$. 

In the previous section we showed that the non-unitary action of the two-qutrit operators $\mathcal{A}^{+}(v)$, $\mathcal{F}^{G}_{\rho_1}$ and $\mathcal{F}^{G}_{\rho_2}$ can be embedded in a larger unitary circuit using three operation ancillas. For a product $\mathcal{M}$ of $n$ such primitives, we denote the corresponding unitary dilation by $M$, chosen such that
\begin{equation}
M\ket{\psi}\ket{0}^{\otimes 3n}_{A}
=
(\mathcal{M}\ket{\psi})\ket{0}^{\otimes 3n}_{A}
+
\ket{\Phi^\perp},
\end{equation}
with $(\mathbf{1}\otimes\bra{0}^{\otimes 3n}_{A})\ket{\Phi^\perp}=0$.
It follows that
\begin{equation}
\bra{\psi}\mathcal{M}\ket{\psi}
=
\bra{0}^{\otimes 3n}_{A}\bra{\psi}
M
\ket{\psi}\ket{0}^{\otimes 3n}_{A}.
\end{equation}
The desired non-unitary expectation value can therefore be extracted by applying a Hadamard test to the unitary dilation $M$. In this construction the operation ancillas are not measured or post-selected during the Hadamard test, as only the Hadamard ancilla is measured.

%In the previous section we have shown that the non-unitary action of the two-qutrit operators $\mathcal{A}^{+}(v), \mathcal{F}^{G}_{\rho_{1}}$ and $\mathcal{F}^{G}_{\rho_{2}}$ may be embedded within a higher-dimensional unitary circuit and implemented probabilistically via post-selection on the ancillary state $\ket{000}_{A}$.
% we have outlined an encoding whereby the non-unitary action of the respective operation is embedded within a higher-dimensional unitary circuit and implemented probabilistically utilising post-selection on a number of ancillas.
%Denoting $\mathcal{M}$ as any $n$-fold product of such operators, this non-unitary operation is thus encoded through some unitary circuit $M$ such that  
% Defining some operator $\mathcal{M}$ as the $n$-fold product of such operators, the non-unitary transformation
% % Any product of these operations 
% $\mathcal{M}:\ket{\psi}\mapsto\ket{\phi}$ is thus encoded as 
%$M:\ket{\psi}\ket{0}^{\otimes 3n}_{A}\mapsto (\mathcal{M}\ket{\psi})\ket{0}^{\otimes 3n}_{A}+\ket{\phi^{\perp}}\ket{0^{\perp}}^{\otimes 3n}$ where $\ket{\phi^{\perp}}$ and $\ket{0^{\perp}}^{\otimes 3n}$ denote states orthogonal to $\ket{\phi}$ and $\ket{0}^{\otimes 3n}$ respectively. Using this notation it thus becomes clear that the desired expectation value $\bra{\psi}\mathcal{M}\ket{\psi}=\bra{0}^{\otimes 3n}_{A}\bra{\psi}M\ket{\psi}\ket{0}^{\otimes 3n}_{A}$.
%As $M$ is just some unitary operation, the desired non-unitary inner product may therefore be reconstructed by performing a Hadamard test on $M$ [refs]. 

Consider first the two-qutrit product
\begin{equation}
    \mathcal{M}_{+} = (U_{Q}^{\dagger}\otimes U_{Q}^{\dagger})\mathcal{F}^{G}_{\rho_{2}}\mathcal{F}^{G}_{\rho_{1}}\mathcal{A}^{+}(v)\mathcal{F}^{G}_{\rho_{2}}\mathcal{F}^{G}_{\rho_{1}}(U_{Q}\otimes U_{Q}),
\end{equation}
such that 
\begin{equation}
    (R^{+})^{2} = \mathcal{N}_{+}\bra{e,e}\mathcal{M}_{+}\ket{e,e},
\end{equation}
with the normalisation $\mathcal{N}_{+}=1/2$. Utilising the construction introduced in the Section \ref{sec:circuits}, $\mathcal{M}_{+}$ is implemented probabilistically through the corresponding unitary circuit $M_{+}$ and a total of 15 ancillas.
%Utilising the method introduced in the previous sections, $\mathcal{M}_{+}$ is implemented probabilistically by post-selecting the result of the corresponding unitary circuit $M_{+}$ on the ancillary state $\ket{0}_{A}^{\otimes 15}$ (see Supplementary Material). When compiled on Quantinuum's processor this circuit uses a total of 21 qubits with 193 single-qubit and 118 two-qubit operations. 
In order to implement a Hadamard test to extract the real component $\text{Re}(\bra{0}^{\otimes 15}_{A}\bra{100,100}M_{+}\ket{100,100}\ket{0}^{\otimes 15}_{A})$, a circuit must thus be constructed that implements the operator $M_{+}$ conditional on the state of an additional Hadamard ancilla \cite{cleve1998quantum}.
% controlled version of the operator $M_{+}$ must be applied. 
Careful inspection of the action  of each gate on this initial state in fact reveals that the total operation $M_{+}$ comprised of 311 total gates, may be effectively conditioned on the Hadamard qubit by employing only 29 additional controlled gates. This full construction is shown on the right-hand side of Figure \ref{fig:Htest}(a). When compiled on the Quantinuum H2 processors, the total circuit used to implement the Hadamard test for $\text{Re}((R^{+})^{2})$ uses a total of 229 single-qubit gates and 134 two-qubit gates acting on 22 qubits. Notably the test requires only a single measurement, on the Hadamard ancilla.
% The final probability distribution was obtained by combining the results of 7000 experimental runs. 
Accounting for normalization factors and the factor of $1/2$ in the implementation of each gate $D_{F}$, the real component of $(R^{+})^{2}$ is thus calculated as 
\begin{equation}
    \text{Re}((R^{+})^{2}) = 8(2p_{0}-1),
    \label{eq:RHad}
\end{equation}
where $p_{0}$ is the probability of measuring `0' on the Hadamard ancilla. 
The imaginary component $\text{Im}((R^{+})^{2})$ is calculated in an identical manner with the additional inclusion of an $S^{\dagger}$ gate after the first Hadamard gate.

% Thus the full sequence of 332 gates becomes effectively controlled on the Hadamard ancilla qubit by implementing the circuit shown in Figure \ref{fig:Htest}, such that just 29 gates are controlled on this qubit

% In general, for some unitary operator $M$, this allows one to calculate -- as the difference between the probability of measuring `0' and `1' on the Hadamard ancilla 
% %for explicitly extracting both the real and imaginary co-efficients. 

% % Non-unitary operations
% % -- Adapted Hadamard test

% We want to find
% \begin{align}
%     (R^{+})^{2} &= \text{Re}(\bra{e,e}\mathcal{M}_{+}\ket{e,e}) \nonumber \\
%     &\qquad\qquad + \text{Im}(\bra{e,e}\mathcal{M}_{+}\ket{e,e}) i,
% \end{align}
% where
% \begin{equation}
%     \mathcal{M}_{+} = (U_{Q}^{\dagger}\otimes U_{Q}^{\dagger})\mathcal{F}^{G}_{\rho_{2}}\mathcal{F}^{G}_{\rho_{1}}\mathcal{A}^{+}(v)\mathcal{F}^{G}_{\rho_{2}}\mathcal{F}^{G}_{\rho_{1}}(U_{Q}\otimes U_{Q}).
% \end{equation}

% Notably, inspection of the -- allows for the 
% full operation to become effectively controlled on the Hadamard ancilla qubit by converting a much smaller number of the

% Thus the full sequence of 332 gates becomes effectively controlled on the Hadamard ancilla qubit by implementing the circuit shown in Figure \ref{fig:Htest}, such that just 29 gates are controlled on this qubit

To determine $\operatorname{Re}[(R^{G})^{2}]$ and $\operatorname{Im}[(R^{G})^{2}]$ with minimal additional resources, we use the resolution of the identity $\mathcal{A}^{+}(v)+\mathcal{A}^{G}(v)=\mathbf{1}_{9}$. Defining $(R^{1})^{2}$ as the auxiliary overlap obtained by replacing $\mathcal{A}^{+}(v)$ with $\mathbf{1}_{9}$ in the $(R^{+})^{2}$ circuit, linearity gives
\begin{equation}
    (R^{G})^{2}=(R^{1})^{2}-(R^{+})^{2}.
\end{equation}
The quantity $(R^{1})^{2}$ is an identity-channel overlap rather than a physical braiding phase.

% \begin{align}
%     (R^{G})^{2} &= \left[ \text{Re}(\bra{e,e}\mathcal{M}_{\mathbf{1}}\ket{e,e})-\text{Re}(\bra{e,e}\mathcal{M}_{+}\ket{e,e}) \right]\nonumber \\
%     &+ \left[ \text{Im}(\bra{e,e}\mathcal{M}_{\mathbf{1}}\ket{e,e})-\text{Im}(\bra{e,e}\mathcal{M}_{+}\ket{e,e}) \right] i, \nonumber
% \end{align}
% where
% \begin{equation}
%     \mathcal{M}_{\mathbf{1}} = (U_{Q}^{\dagger}\otimes U_{Q}^{\dagger})\mathcal{F}^{G}_{\rho_{2}}\mathcal{F}^{G}_{\rho_{1}}\mathbf{1}_{9}\mathcal{F}^{G}_{\rho_{2}}\mathcal{F}^{G}_{\rho_{1}}(U_{Q}\otimes U_{Q}).
% \end{equation}
As the two-qutrit operation $\mathcal{M}_{\mathbf{1}}$ has no projection operator, the controlled version of the corresponding unitary encoding $M_{\mathbf{1}}$ uses only 12 ancillary qubits.
%19 qubits.
%with 223 single-qubit gates and 128 two-qubit gates.

The Hadamard tests to determine $\text{Re}(\bra{0}^{\otimes 12}_{A}\bra{100,100}M_{\mathbf{1}}\ket{100,100}\ket{0}^{\otimes 12}_{A})$ and $\text{Im}(\bra{0}^{\otimes 12}_{A}\bra{100,100}M_{\mathbf{1}}\ket{100,100}\ket{0}^{\otimes 12}_{A})$ as the difference in measurement probabilities on the Hadamard qubit are implemented as for $R^{+}$. The real component circuit requires 223 single-qubit and 128 two-qubit gates, while the imaginary component requires an additional single-qubit $S^{\dagger}$ on the Hadamard qubit. 
% {\color{red} [Why number of gates is different? Some correspond to measurements?]}

% Implementing eq \ref{eq:R2} used 22 qubits and 385 total gates, 229 single qubit gates and 134 two qubit gates.
% Implementing eq \ref{identity} used 19 qubits and 370 total gates, comprising 223 single qubit gates and 128 two qubit gates. As these were Hadamard tests these measurements therefore required only one measurement gate.
% The final probability distribution was obtained by combining the results of 7000 experimental runs. 

% As these were Hadamard tests these measurements therefore required only one measurement gate.
% The final probability distribution was obtained by combining the results of 13000 experimental runs. 

The number of shots was chosen to give a target binomial uncertainty on the extracted quadratures. 
For $\text{Re}((R^{+})^{2})$ and $\text{Im}((R^{+})^{2})$ as calculated in Eq.~\eqref{eq:RHad}, for example, the corresponding uncertainties are expressed as functions of the uncertainty in the measured probability $p_{0}$.
For a binomial distribution
\begin{equation}
    \text{Var}(p_{0}) = \frac{p_{0}(1-p_{0})}{N},
\end{equation}
where $N$ is the total number of shots. Propagating this through Eq.~\eqref{eq:RHad}, we obtain
\begin{equation}
    \sigma_{R^{+}} = \left|\frac{dRe(R^{+})}{dp_{0}}\right|\sqrt{\text{Var}(p_{0})} = 16\sqrt{\frac{p_{0}(1-p_{0})}{N}}.
\label{eqn:bin}
\end{equation}

We performed a total of $N=7000$ shots yielding the measured ancilla probability $p_0$ and the corresponding reconstructed values were
\begin{equation}
\begin{array}{lcc}
\hline
\text{Quantity} & p_0 & \text{Extracted value} \\
\hline
\mathrm{Re}(R^+)^2 & 0.4700 & -0.480\pm0.095 \\
\mathrm{Im}(R^+)^2 & 0.4467 & -0.853\pm0.095 \\
\mathrm{Re}(R^{\mathbf 1})^2 & 0.4439 & -0.898\pm0.095 \\
\mathrm{Im}(R^{\mathbf 1})^2 & 0.4996 & -0.007\pm0.095 \\
\hline
\end{array}
\end{equation}
This results in the value $\text{Re}((R^{+})^{2})=-0.48$. Substituting into Eq.~\eqref{eqn:bin}, we thus estimate the binomial uncertainty in this value 
%this experimental configuration corresponds to an uncertainty 
as 
\begin{equation}
    \sigma_{\text{Re}(R^{+})}=0.095.
\end{equation}

For the case of `failure' within this system, we consider the complete depolarization of the Hadamard ancilla such that the measurement of `0' or `1' becomes random, yielding $p_{0}^{\text{failure}}=1/2$ and correspondingly $(R^{+})^{2, \text{failure}}=0$.
Our analysis thus shows that $N=7000$ is sufficient to generate a value for $\text{Re}((R^{+})^{2})$ that is distinct from the case of failure by over $5 \sigma_{\text{Re}(R^{+})}$.

% elements are 
% As both $Re(R^{+})$ and $Im(R^{+})$ are calculated as functions of the probability $p_{0}$ in their respective circuits, the uncertainty 

% Assume binomial sampling statistics $\rightarrow$ error bars come from propagating shot noise through calculation of matrix elements.

% Both $Re(R^{+})$ and $Im(R^{+})$ are calculated using the same expression:
% \begin{equation}
%     Re(R^{+}) = 8(p_{0}-p_{1}) = 8(2p_{0}-1),
%     \label{eq:R+}
% \end{equation}
% So the uncertainty in the elements $Re(R^{+})$ and $Im(R^{+})$ is entirely determined by the uncertainty in $p_{0}$.

% For a binomial distribution
% \begin{equation}
%     \text{Var}(p_{0}) = \frac{p_{0}(1-p_{0})}{N},
% \end{equation}
% propagating this through Eq.~\eqref{eq:R+},
% \begin{equation}
%     \sigma_{R^{+}} = \left|\frac{dRe(R^{+})}{dp_{0}}\right|\sqrt{\text{Var}(p_{0})} = 16\sqrt{\frac{p_{0}(1-p_{0})}{N}}.
% \end{equation}

% For $Re(R^{+})$ we have $N=7000$ yielding the experimental value $Re(R^{+})=-0.480$ $\rightarrow$ $p_{0}=0.47$. Substituting in these values gives
% \begin{equation}
%     \sigma_{Re(R^{+})} = 0.095
% \end{equation}
% such that
% \begin{equation}
%     Re(R^{+}) = -0.480 \pm 0.095.
% \end{equation}

% Similarly for $Im(R^{+})$, from $N=7000$ we have $Im(R^{+})=-0.853$ $\rightarrow$ $p_{0}=0.447$ such that
% \begin{equation}
%     \sigma_{Im(R^{+})} = 0.095
% \end{equation}
% again and
% \begin{equation}
%     Im(R^{+}) = -0.853 \pm 0.095,
% \end{equation}
% as shown in Figure [].

In our protocol the value $R^{G}$ is calculated as
\begin{equation}
    (R^{G})^2 = (R^{\mathbf{1}})^2 - (R^{+})^2,
\end{equation}
therefore for both the real and imaginary components of $R^{G}$ we must add the uncertainties from $R^{\mathbf{1}}$ and $R^{+}$ in quadrature as
\begin{equation}
    \sigma_{R^{G}} = \sqrt{\sigma_{R^{\mathbf{1}}}^{2}+\sigma_{R^{+}}^{2}}.
\end{equation}
The results of this calculation are shown in Sec.~\ref{sec:R}.

\subsubsection{Experimental reconstruction of the fusion amplitudes}

Having described the reconstruction of the squared braiding phases on Quantinuum hardware, we now turn to the experimental extraction of the fusion amplitudes. As discussed in Eqs.~\eqref{eq:F21} and \eqref{eq:F22}, the squared fusion matrix elements $(F^{i}_{j})^{2}$, with $i,j\in\{A,B,G\}$, can be obtained from overlaps of ribbon and charge-projection operators~\cite{Byles2024}. In the reduced two-qutrit protocol, only the sectors $i,j\in\{+,G\}$, with $+=A+B$, are directly accessible. The relevant two-qutrit overlap amplitudes are
\begin{equation}
     m_{ji}
     =
     \bra{\xi}\mathcal{O}_{m_{ji}}\ket{\xi},
     \qquad
     i,j\in\{+,G\},
\end{equation}
where
\begin{equation}
    \mathcal{O}_{m_{ji}}
    =
    \mathcal{F}^{G}_{\rho_{2}}
    \mathcal{F}^{G}_{\rho_{2}}
    \mathcal{A}^{j}(v)
    \mathcal{F}^{G}_{\rho_{1}}
    \mathcal{A}^{G}(v)
    \mathcal{F}^{G}_{\rho_{2}}
    \mathcal{A}^{i}(v)
    \mathcal{F}^{G}_{\rho_{2}}
    \mathcal{F}^{G}_{\rho_{1}} .
    \label{eq:mji}
\end{equation}
The corresponding normalisation factors are
\begin{equation}
     n_{j}
     =
     \bra{\xi}\mathcal{O}_{n_{j}}\ket{\xi},
     \qquad
     j\in\{+,G\},
\end{equation}
with
\begin{equation}
    \mathcal{O}_{n_{j}}
    =
    \mathcal{F}^{G}_{\rho_{2}}
    \mathcal{F}^{G}_{\rho_{2}}
    \mathcal{A}^{j}(v)
    \mathcal{F}^{G}_{\rho_{1}}
    \mathcal{A}^{G}(v)
    \mathcal{F}^{G}_{\rho_{1}}
    \mathcal{A}^{j}(v)
    \mathcal{F}^{G}_{\rho_{2}}
    \mathcal{F}^{G}_{\rho_{2}} .
    \label{eq:nj}
\end{equation}

These quantities determine the squared fusion matrix
\begin{equation}
    \Phi_{ij}\equiv |F_{ij}|^{2}
    =
    \begin{pmatrix}
        a & a & b\\
        a & a & b\\
        b & b & c
    \end{pmatrix},
\end{equation}
which, for the ideal $D(S_3)$ data, is
\begin{equation}
    \Phi
    =
    \frac{1}{4}
    \begin{pmatrix}
      1 & 1 & 2\\
      1 & 1 & 2\\
      2 & 2 & 0
    \end{pmatrix}.
\end{equation}
At the amplitude level the independent entries are related to the overlaps by
\begin{equation}
     a = \frac{m_{++}}{2n_{+}},
     \,\,
     b = \frac{1}{2}\left(\frac{m_{+G}}{n_{+}}+\frac{m_{G+}}{2\omega n_{G}}\right), 
     \,\,
     c = \frac{m_{GG}}{\omega n_{G}}.
     \label{eq:F-amplitude-ratios}
\end{equation}
% The lower-right entry is zero because $F^{G}_{GGG}$ has no $G\to G$ matrix element in this basis.

Experimentally, the circuits established to determine the amplitudes $m_{ji}$ and $n_j$  return post-selected probabilities associated with the corresponding overlap and normalisation events, which we denote by $p(m_{ji})$ and $p(n_j)$. For the self-dual $D(S_3)$ model considered here, the relevant $F$-symbols can be chosen real, as shown in Refs.~\cite{MattPeterJJiannisToAppear1,MattPeterJJiannisToAppear2}. We therefore do not need to reconstruct additional phase information for these fusion elements. Moreover, the charge-neutral ribbon geometry contributes the conjugate fusion amplitude at the opposite ribbon endpoint, so the measured closed diagram gives $F_{ij}F_{ij}^{*}=|F_{ij}|^2$. The independent entries are therefore reconstructed directly from probability ratios as
% \begin{align}
%     &a=
%     \frac{1}{2}
%     \sqrt{\frac{p(m_{++})}{p(n_+)}},
%     \qquad
%     b=
%     \sqrt{\frac{p(m_{+G})}{p(n_+)}},\nonumber\\
%     &c=
%     \frac{1}{2}
%     \sqrt{\frac{p(m_{G+})}{p(n_G)}}, \qquad d=\sqrt{\frac{p(m_{GG}}{p(n_{G})}}.
%      \label{eq:F-probability-ratios}
% \end{align}
\begin{align}
    &a=
    \frac{1}{2}
    \sqrt{\frac{p(m_{++})}{p(n_+)}}, \nonumber \\
    &b= \frac{1}{2}\left(\sqrt{\frac{p(m_{+G})}{p(n_+)}}+
    \frac{1}{2}
    \sqrt{\frac{p(m_{G+})}{p(n_G)}}\right), \label{eq:F-probability-ratios}\\
    &
    c= \sqrt{\frac{p(m_{GG})}{p(n_{G})}}. \nonumber
\end{align}
Thus, unlike the braiding phases, the fusion amplitudes do not require a Hadamard test. They are reconstructed from post-selected probabilities, at the cost of a reduced effective sample size.

To obtain the probabilities $p(m_{ji})$, we implement the circuit
\begin{equation}
    (U_{\mathrm{Q}}^{\dagger}\otimes U_{\mathrm{Q}}^{\dagger})
    \mathcal{O}_{m_{ji}}
    (U_{\mathrm{Q}}\otimes U_{\mathrm{Q}})
    \label{eq:circ_mji}
\end{equation}
with
\begin{equation}
    \mathcal{O}_{m_{ji}}
    =
    \mathcal{F}^{G}_{\rho_{2}}
    \mathcal{F}^{G}_{\rho_{2}}
    \mathcal{A}^{j}(v)
    \mathcal{F}^{G}_{\rho_{1}}
    \mathcal{A}^{G}(v)
    \mathcal{F}^{G}_{\rho_{2}}
    \mathcal{A}^{i}(v)
    \mathcal{F}^{G}_{\rho_{2}}
    \mathcal{F}^{G}_{\rho_{1}},
    \label{eq:Mmji}
\end{equation}
on the initial encoded state
\begin{equation}
    \ket{e,e}
    \equiv
    \ket{1,0,0}\ket{1,0,0}.
\end{equation}
Using the coherent implementation of the charge projector described in Sec.~\ref{sec:proj}, the same circuit identifies the different choices $i,j\in\{+,G\}$ through the projection-ancilla readouts. Successful implementation of each ribbon operator is heralded by measuring the corresponding operation ancillas in $\ket{0}$. The charge outcome $+$ or $G$ is then identified by the projection register, with the syndromes described in Eq.~\eqref{eq:Acases}. Finally, selecting the branch in which the computational qubits return to $\ket{e,e}$ yields the post-selected probabilities $p(m_{ji})$.

% Importantly, utilising the second form of the projection operator introduced in Section \ref{sec:proj}, this circuit simultaneously implements all combinations $\mathcal{O}_{m_{ji}},$ $i,j\in\{+,G\}$, with the outcome of each specific operation being uniquely reconstructed from the correct combination of ancilla readouts. 
% For the combination of ribbon and projection operators as given in Eq.~\eqref{eq:mji} \textcolor{purple}{there are a total of 27 ancillary qubits}. Successful implementation of each ribbon operator is heralded by a measurement of `0' on each operation ancilla, while the projection operator $\mathcal{A}^{i}(v)$ implemented 
% as in Eq.~\eqref{eq:Acases}, uniquely identifies each combination $i=+$ or $i=G$ by the readout `$00$' or `$11$' on the Toffoli ancillas respectively and `0' on all others. 

% The projection operator $\mathcal{A}^{i}(v)$ is then implemented 
% as in Eq.~\eqref{eq:Acases}, such that each combination of $i=+$ or $i=G$ may be separately identified by the readout `$00$' or `$11$' on the Toffoli ancillas respectively and `0' on all others.
% Finally, selecting the unique outcome for which the computational qubits are in the state $\ket{1,0,0}\ket{1,0,0}$ thus allows for direct read-out of the probability

This circuit uses a total of 36 qubits: 30 ancillas and 6 computational qubits. Here we note the identity $\left(\mathcal{F}^{G}_{\rho_2}\right)^{2}=
(U_{\mathrm{Q}}^{\dagger}\otimes \mathbf{1}_3) \widetilde{\mathcal{D}}_{F}(U_{\mathrm{Q}}\otimes \mathbf{1}_3)$, such that repeated application of the ribbon operator may be implemented with a single non-unitary component $\widetilde{\mathcal{D}}_{F}=(\mathcal{D}_{F})^{2}$. This operation is
% the non-unitary action of the squared operation is captured by the single gate $\widetilde{\mathcal{D}}_{F}=(\mathcal{D}_{F})^{2}$, 
% repeated application of two $\mathcal{F}^{G}_{\rho_{2}}$ ribbon operators may be enacted
% has been used to reduce the number of required ancillas. As $\widetilde{\mathcal{D}}_{F}=(\mathcal{D}_{F})^{2}$, 
 simply constructed as in Fig.~\ref{fig:app-ribbon-operators}(b) with $d_{F}(\theta)=d_{F}(\arccos{1/4})$. The number of ancillas is thus reduced by three
 % reducing the number of required ancillas by three 
 compared to direct re-application of $\mathcal{F}^{G}_{\rho_{2}}$. 
Within this simplified circuit there are a total of 322 single-qubit gates, 213 two-qubit gates and 36 measurements.
%provides a reduction in the number of ancillas, as this amounts to removing one

% \textcolor{gray}{[Insert here number of single-qubit gates, number of two-qubit gates, circuit depth, backend used, and number of shots for the $m_{ji}$ circuit.]}

The normalisation probabilities are obtained similarly from a second circuit implementing
\begin{equation}
    (U_{\mathrm{Q}}^{\dagger}\otimes U_{\mathrm{Q}}^{\dagger})
    \mathcal{O}_{n_j}
    (U_{\mathrm{Q}}\otimes U_{\mathrm{Q}}),
    \label{eq:circ_nj}
\end{equation}
where
\begin{equation}
    \mathcal{O}_{n_{j}}
    =
    \mathcal{F}^{G}_{\rho_{2}}
    \mathcal{F}^{G}_{\rho_{2}}
    \mathcal{A}^{j}(v)
    \mathcal{F}^{G}_{\rho_{1}}
    \mathcal{A}^{G}(v)
    \mathcal{F}^{G}_{\rho_{1}}
    \mathcal{A}^{j}(v)
    \mathcal{F}^{G}_{\rho_{2}}
    \mathcal{F}^{G}_{\rho_{2}}.
    \label{eq:Mnj}
\end{equation}
The two possible normalisation channels, $n_+$ and $n_G$, are again distinguished by the projection-ancilla readouts, while successful ribbon operations are heralded by the corresponding operation ancillas. Post-selecting on the computational output $\ket{e,e}$ gives the probabilities $p(n_+)$ and $p(n_G)$. As the operation $\mathcal{O}_{n_{j}}$ contains two instances of $(\mathcal{F}^{G}_{\rho_{2}})^{2}$ the number of ancillas is further reduced for a total of 33 qubits. This circuit is thus implemented using 275 single-qubit gates, 184 two-qubit gates and 33 measurements.

% \textcolor{gray}{[Insert here number of computational qubits, number of operation ancillas, number of projection ancillas, total number of qubits for the $n_j$ circuit.]}

% \textcolor{gray}{[Insert here number of single-qubit gates, number of two-qubit gates, circuit depth, backend used, and number of shots for the $n_j$ circuit.]}

The uncertainties in $a$ and $b$ are obtained by assuming binomial sampling statistics for the post-selected probabilities and propagating these uncertainties through Eq.~\eqref{eq:F-probability-ratios}. 
Assume the circuit implementing Eq.~\eqref{eq:circ_mji} runs for $N_{1}$ shots and the second for Eq.~\eqref{eq:circ_nj} for $N_{2}$ shots. The uncertainties on each of the corresponding measured probabilities are calculated as
\begin{equation}
    \begin{split}
        \sigma_{p(m_{ji})} = \sqrt{\frac{p( m_{ji})(1-p( m_{ji}))}{N_{1}}}, \\
        \sigma_{p(n_{j})} = \sqrt{\frac{p(n_{j})(1-p(n_{j}))}{N_{2}}}.
    \end{split}
\end{equation}
% \begin{equation}
%     \begin{split}
%         \sigma_{p({m}_{++})} = \sqrt{\frac{p({m}_{++})(1-p({m}_{++}))}{N_{1}}}, \\
%         \sigma_{p( {m}_{+G})} = \sqrt{\frac{p( {m}_{+G})(1-p( {m}_{+G}))}{N_{1}}}, \\
%         \sigma_{p( {m}_{G+})} = \sqrt{\frac{p( {m}_{G+})(1-p( {m}_{G+}))}{N_{1}}}
%     \end{split}  
% \end{equation}
% and
% \begin{equation}
%     \sigma_{p(n_{j})} = \sqrt{\frac{p(n_{j})(1-p(n_{j}))}{N_{2}}}.
% \end{equation}
% \begin{equation}
%     \begin{split}
%         \sigma_{p( {n}_{+})} = \frac{p( {n}_{+})(1-p( {n}_{+}))}{N_{2}}, \\
%         \sigma_{p( {n}_{G})} = \frac{p( {n}_{G})(1-p( {n}_{G}))}{N_{2}}.
%     \end{split}
% \end{equation}
 In order to calculate the uncertainties on the matrix elements $a,b$ and $c$, we must then propagate the uncertainties on each measurement through Equation \eqref{eq:F-probability-ratios}.
 Using logarithmic differentiation the relative uncertainty is found to be
 \begin{equation}
     \left(\frac{\sigma_{a}}{a}\right)^{2} = \frac{1}{4} \left[ \left(\frac{\sigma_{p(  m_{++})}}{p(  m_{++})}\right)^{2}+ \left(\frac{\sigma_{p(  n_{+})}}{p(  n_{+})}\right)^{2}\right],
 \end{equation}
yielding
\begin{equation}
    \sigma_{a} = \frac{a}{2}\sqrt{\frac{1-p(  m_{++})}{p(  m_{++})N_{1}} + \frac{1-p(  n_{+})}{p(  n_{+})N_{2}}}.
\end{equation}
% \begin{align}
%      \sigma_{a} &= \frac{a}{2}\sqrt{\left(\frac{\sigma_{p( {m}_{++})}}{p( {m}_{++})}\right)^{2}+ \left(\frac{\sigma_{p( {n}_{+})}}{p( {n}_{+})}\right)^{2}}, \\
%      &= \frac{a}{2}\sqrt{\frac{1-p( {m}_{++})}{p( {m}_{++})N_{1}} + \frac{1-p( {n}_{+})}{p( {n}_{+})N_{2}}}.
% \end{align}
% Similarly,
% \begin{equation}
%     \sigma_{b} = \frac{b}{2}\sqrt{\frac{1-p( {m}_{+G})}{p( {m}_{+G})N_{1}} + \frac{1-p( {n}_{+})}{p( {n}_{+})N_{2}}},
% \end{equation}
% and
% \begin{equation}
%     \sigma_{c} = \frac{c}{2}\sqrt{\frac{1-p( {m}_{G+})}{p( {m}_{G+})N_{1}} + \frac{1-p( {n}_{G})}{p( {n}_{G})N_{2}}}.
% \end{equation}
The expression for the uncertainty $\sigma_{b}$ is found analogously. For $N_{1}= 11,000$, $N_{2}=13,000$, these post-selected probabilities are given by:
\[
\begin{array}{cccccc}
\hline
\text{Post-selected } & \text{Accepted counts} & \text{Measured value} \\
\text{probability} & & \\
\hline
p(  m_{++}) & 37 & 0.00336\pm0.00055\\
p(  m_{+G}) & 30 & 0.00273\pm0.00050\\
p(  m_{G+}) & 11 & 0.00100 \pm 0.00030\\
p( m_{GG}) & 0 & 0.00000 \\
p(  n_{+}) & 179 & 0.01377\pm 0.00102\\
p(  n_{G}) & 6 & (4.6\pm1.9)\times10^{-4}\\
\hline
\end{array}
\]
% {\color{red} [Should it be $179/15000 = 0.0119$ rather yhan 0.0133? Also $6/15000=4 \times 10^{-4}$ rather than $2.2 \times 10^{-4}$? Also from these values I get $c=1.02...$ not $0.708$. Can you check {\bf all} values again with the provided table after you correct/check the table values?]}
obtained by conditioning on the required ribbon-ancilla, projection-ancilla and computational-output syndromes. No accepted events were observed in the $m_{GG}$ channel, thus reproducing $c=0$ as is consistent with the theoretical value. 
%In reconstructing $F_{\mathrm{exp}}$, we impose the exact $D(S_3)$ condition $F_{GG}=0$, and hence $c=\Phi_{GG}=0$. The zero observed count is consistent with this condition but does not establish it independently.

For the braiding phases, the chosen shot numbers were sufficient to resolve the experimentally obtained values with more than five-standard-deviation separation from the zero signal expected under complete depolarisation. For the fusion reconstruction, the limiting channel is instead the small normalisation probability $p(  n_G)=4.6\times10^{-4}$. With $N_2=13,000$ shots this corresponds to only $6$ accepted events. Reducing the contribution of this channel to the relative uncertainty in $b$ to the 10\% level would require of order $10^5$ normalisation shots, while obtaining a few hundred accepted $n_G$ events would require close to $10^6$ shots. The precision of the present fusion reconstruction is therefore limited
primarily by the small number of accepted post-selected events in the
$G$-normalisation channel.

\acknowledgements 

We would like to thank Ewan Forbes and Ryan Smith for useful conversations. This work was supported by EPSRC with Grant Nos. EP/W524372/1 and
UKRI1337:Anyons24.

\section*{Data availability}

All data supporting the findings of this study, including the raw measurement counts, processed data and source data underlying the figures, are publicly available, together with the analysis code, 
% in the ``Reconstructing non-Abelian braiding and fusion without anyon transport'' repository 
via Zenodo at \href{https://doi.org/10.5281/zenodo.21508462}{https://doi.org/10.5281/zenodo.21508462}.

\section*{Author contributions}

J.K.P. and L.B. carried out the theoretical analysis. M.D.H. developed the circuit decompositions for the required operations. B.T.H.V. implemented the operations in code, and B.T.H.V. and L.B. performed the experiments on the quantum hardware. All authors contributed to writing and revising the manuscript.

% \section*{Competing interests}

% The authors declare no competing interests.

%\bibliographystyle{ieeetr}

% \bibliography{refs}

\begin{thebibliography}{10}
\expandafter\ifx\csname url\endcsname\relax
  \def\url#1{\texttt{#1}}\fi
\expandafter\ifx\csname urlprefix\endcsname\relax\def\urlprefix{URL }\fi
\providecommand{\bibinfo}[2]{#2}
\providecommand{\eprint}[2][]{\url{#2}}

\bibitem{Kitaev2003}
\bibinfo{author}{Kitaev, A.~Y.}
\newblock \bibinfo{title}{Fault-tolerant quantum computation by anyons}.
\newblock \emph{\bibinfo{journal}{Annals of Physics}} \textbf{\bibinfo{volume}{303}}, \bibinfo{pages}{2--30} (\bibinfo{year}{2003}).

\bibitem{Nayak2008}
\bibinfo{author}{Nayak, C.}, \bibinfo{author}{Simon, S.~H.}, \bibinfo{author}{Stern, A.}, \bibinfo{author}{Freedman, M.} \& \bibinfo{author}{Das~Sarma, S.}
\newblock \bibinfo{title}{Non-abelian anyons and topological quantum computation}.
\newblock \emph{\bibinfo{journal}{Reviews of Modern Physics}} \textbf{\bibinfo{volume}{80}}, \bibinfo{pages}{1083--1159} (\bibinfo{year}{2008}).

\bibitem{Freedman2003}
\bibinfo{author}{Freedman, M.~H.}, \bibinfo{author}{Kitaev, A.}, \bibinfo{author}{Larsen, M.~J.} \& \bibinfo{author}{Wang, Z.}
\newblock \bibinfo{title}{Topological quantum computation}.
\newblock \emph{\bibinfo{journal}{Bulletin of the American Mathematical Society}} \textbf{\bibinfo{volume}{40}}, \bibinfo{pages}{31--38} (\bibinfo{year}{2003}).

\bibitem{Bonderson2008PRL}
\bibinfo{author}{Bonderson, P.}, \bibinfo{author}{Freedman, M.} \& \bibinfo{author}{Nayak, C.}
\newblock \bibinfo{title}{Measurement-only topological quantum computation}.
\newblock \emph{\bibinfo{journal}{Physical Review Letters}} \textbf{\bibinfo{volume}{101}}, \bibinfo{pages}{010501} (\bibinfo{year}{2008}).

\bibitem{Bonderson2009AnnPhys}
\bibinfo{author}{Bonderson, P.}, \bibinfo{author}{Freedman, M.} \& \bibinfo{author}{Nayak, C.}
\newblock \bibinfo{title}{Measurement-only topological quantum computation via anyonic interferometry}.
\newblock \emph{\bibinfo{journal}{Annals of Physics}} \textbf{\bibinfo{volume}{324}}, \bibinfo{pages}{787--826} (\bibinfo{year}{2009}).

\bibitem{BravyiKimKlieschKoenig2022Adaptive}
\bibinfo{author}{Bravyi, S.}, \bibinfo{author}{Kim, I.}, \bibinfo{author}{Kliesch, A.} \& \bibinfo{author}{Koenig, R.}
\newblock \bibinfo{title}{Adaptive constant-depth circuits for manipulating non-abelian anyons} (\bibinfo{year}{2022}).
\newblock \eprint{2205.01933}.

\bibitem{TantivasadakarnVishwanathVerresen2023Hierarchy}
\bibinfo{author}{Tantivasadakarn, N.}, \bibinfo{author}{Vishwanath, A.} \& \bibinfo{author}{Verresen, R.}
\newblock \bibinfo{title}{Hierarchy of topological order from finite-depth unitaries, measurement, and feedforward}.
\newblock \emph{\bibinfo{journal}{PRX Quantum}} \textbf{\bibinfo{volume}{4}}, \bibinfo{pages}{020339} (\bibinfo{year}{2023}).
\newblock \eprint{2209.06202}.

\bibitem{VerresenTantivasadakarnVishwanath2021Efficiently}
\bibinfo{author}{Verresen, R.}, \bibinfo{author}{Tantivasadakarn, N.} \& \bibinfo{author}{Vishwanath, A.}
\newblock \bibinfo{title}{Efficiently preparing schr{\"o}dinger's cat, fractons and non-abelian topological order in quantum devices} (\bibinfo{year}{2021}).
\newblock \eprint{2112.03061}.

\bibitem{RenTantivasadakarnWilliamson2025SolvableAnyons}
\bibinfo{author}{Ren, Y.}, \bibinfo{author}{Tantivasadakarn, N.} \& \bibinfo{author}{Williamson, D.~J.}
\newblock \bibinfo{title}{Efficient preparation of solvable anyons with adaptive quantum circuits}.
\newblock \emph{\bibinfo{journal}{Physical Review X}} \textbf{\bibinfo{volume}{15}}, \bibinfo{pages}{031060} (\bibinfo{year}{2025}).
\newblock \eprint{2411.04985}.

\bibitem{LevinWen2005StringNet}
\bibinfo{author}{Levin, M.~A.} \& \bibinfo{author}{Wen, X.-G.}
\newblock \bibinfo{title}{String-net condensation: A physical mechanism for topological phases}.
\newblock \emph{\bibinfo{journal}{Physical Review B}} \textbf{\bibinfo{volume}{71}}, \bibinfo{pages}{045110} (\bibinfo{year}{2005}).
\newblock \eprint{cond-mat/0404617}.

\bibitem{BombinMartinDelgado2008NonAbelianKitaev}
\bibinfo{author}{Bombin, H.} \& \bibinfo{author}{Martin-Delgado, M.~A.}
\newblock \bibinfo{title}{Family of non-abelian kitaev models on a lattice: Topological condensation and confinement}.
\newblock \emph{\bibinfo{journal}{Physical Review B}} \textbf{\bibinfo{volume}{78}}, \bibinfo{pages}{115421} (\bibinfo{year}{2008}).
\newblock \eprint{0712.0190}.

\bibitem{BeigiShorWhalen2011QuantumDoubleBoundary}
\bibinfo{author}{Beigi, S.}, \bibinfo{author}{Shor, P.~W.} \& \bibinfo{author}{Whalen, D.}
\newblock \bibinfo{title}{The quantum double model with boundary: Condensations and symmetries}.
\newblock \emph{\bibinfo{journal}{Communications in Mathematical Physics}} \textbf{\bibinfo{volume}{306}}, \bibinfo{pages}{663--694} (\bibinfo{year}{2011}).
\newblock \eprint{1006.5479}.

\bibitem{OgburnPreskill1999TopologicalQC}
\bibinfo{author}{Ogburn, R.~W.} \& \bibinfo{author}{Preskill, J.}
\newblock \bibinfo{title}{Topological quantum computation}.
\newblock In \emph{\bibinfo{booktitle}{Quantum Computing and Quantum Communications}}, vol. \bibinfo{volume}{1509} of \emph{\bibinfo{series}{Lecture Notes in Computer Science}}, \bibinfo{pages}{341--356} (\bibinfo{publisher}{Springer}, \bibinfo{year}{1999}).

\bibitem{Mochon2003NonsolvableAnyons}
\bibinfo{author}{Mochon, C.}
\newblock \bibinfo{title}{Anyons from nonsolvable finite groups are sufficient for universal quantum computation}.
\newblock \emph{\bibinfo{journal}{Physical Review A}} \textbf{\bibinfo{volume}{67}}, \bibinfo{pages}{022315} (\bibinfo{year}{2003}).
\newblock \eprint{quant-ph/0206128}.

\bibitem{Mochon2004SmallerGroups}
\bibinfo{author}{Mochon, C.}
\newblock \bibinfo{title}{Anyon computers with smaller groups}.
\newblock \emph{\bibinfo{journal}{Physical Review A}} \textbf{\bibinfo{volume}{69}}, \bibinfo{pages}{032306} (\bibinfo{year}{2004}).
\newblock \eprint{quant-ph/0306063}.

\bibitem{CuiHongWang2015}
\bibinfo{author}{Cui, S.~X.}, \bibinfo{author}{Hong, S.-M.} \& \bibinfo{author}{Wang, Z.}
\newblock \bibinfo{title}{Universal quantum computation with weakly integral anyons}.
\newblock \emph{\bibinfo{journal}{Quantum Information Processing}} \textbf{\bibinfo{volume}{14}}, \bibinfo{pages}{2687--2727} (\bibinfo{year}{2015}).

\bibitem{ChenRen2025}
\bibinfo{author}{Chen, L.} \& \bibinfo{author}{Ren, T.}
\newblock \bibinfo{title}{A universal circuit set using the $s_3$ quantum double}.
\newblock \emph{\bibinfo{journal}{npj Quantum Information}} \textbf{\bibinfo{volume}{11}}, \bibinfo{pages}{13} (\bibinfo{year}{2025}).

\bibitem{BravyiKitaev2005}
\bibinfo{author}{Bravyi, S.} \& \bibinfo{author}{Kitaev, A.}
\newblock \bibinfo{title}{Universal quantum computation with ideal clifford gates and noisy ancillas}.
\newblock \emph{\bibinfo{journal}{Physical Review A}} \textbf{\bibinfo{volume}{71}}, \bibinfo{pages}{022316} (\bibinfo{year}{2005}).

\bibitem{Byles2024}
\bibinfo{author}{Byles, L.}, \bibinfo{author}{Forbes, E.} \& \bibinfo{author}{Pachos, J.~K.}
\newblock \bibinfo{title}{Demonstration of magic state power of $d(s_3)$ anyons with two qudits}.
\newblock \emph{\bibinfo{journal}{New Journal of Physics}} \textbf{\bibinfo{volume}{28}}, \bibinfo{pages}{044501} (\bibinfo{year}{2026}).

\bibitem{Andersen2023}
\bibinfo{author}{Andersen, T.~I.} \emph{et~al.}
\newblock \bibinfo{title}{Non-abelian braiding of graph vertices in a superconducting processor}.
\newblock \emph{\bibinfo{journal}{Nature}} \textbf{\bibinfo{volume}{618}}, \bibinfo{pages}{264--269} (\bibinfo{year}{2023}).

\bibitem{Iqbal2024}
\bibinfo{author}{Iqbal, M.}, \bibinfo{author}{Tantivasadakarn, N.}, \bibinfo{author}{Verresen, R.} \emph{et~al.}
\newblock \bibinfo{title}{Non-abelian topological order and anyons on a trapped-ion processor}.
\newblock \emph{\bibinfo{journal}{Nature}} \textbf{\bibinfo{volume}{626}}, \bibinfo{pages}{505--511} (\bibinfo{year}{2024}).

\bibitem{Xu2024}
\bibinfo{author}{Xu, S.}, \bibinfo{author}{Sun, Z.-Z.}, \bibinfo{author}{Wang, K.} \emph{et~al.}
\newblock \bibinfo{title}{Non-abelian braiding of fibonacci anyons with a superconducting processor}.
\newblock \emph{\bibinfo{journal}{Nature Physics}} \textbf{\bibinfo{volume}{20}}, \bibinfo{pages}{1469--1475} (\bibinfo{year}{2024}).

\bibitem{Goel2024}
\bibinfo{author}{Goel, S.} \emph{et~al.}
\newblock \bibinfo{title}{Unveiling the non-abelian statistics of $d(s_3)$ anyons using a classical photonic simulator}.
\newblock \emph{\bibinfo{journal}{Physical Review Letters}} \textbf{\bibinfo{volume}{132}}, \bibinfo{pages}{110601} (\bibinfo{year}{2024}).

\bibitem{Lo2026}
\bibinfo{author}{Lo, C. F.~B.} \emph{et~al.}
\newblock \bibinfo{title}{Universal gates from braiding and fusing anyons on quantum hardware}.
\newblock \emph{\bibinfo{journal}{Nature}} \textbf{\bibinfo{volume}{655}}, \bibinfo{pages}{591--597} (\bibinfo{year}{2026}).

\bibitem{Moses2023RaceTrackH2}
\bibinfo{author}{Moses, S.~A.} \emph{et~al.}
\newblock \bibinfo{title}{A race-track trapped-ion quantum processor}.
\newblock \emph{\bibinfo{journal}{Physical Review X}} \textbf{\bibinfo{volume}{13}}, \bibinfo{pages}{041052} (\bibinfo{year}{2023}).
\newblock \eprint{2305.03828}.

\bibitem{MattPeterJJiannisToAppear1}
\bibinfo{author}{Buican, M.}, \bibinfo{author}{Huston, P.} \& \bibinfo{author}{Pachos, J.~K.}
\newblock \bibinfo{title}{Anyons and inherently complex {$F$}-symbols} (\bibinfo{year}{2026}).
\newblock \eprint{2607.10181}.

\bibitem{LeoneOlivieroHamma2022StabilizerRenyi}
\bibinfo{author}{Leone, L.}, \bibinfo{author}{Oliviero, S. F.~E.} \& \bibinfo{author}{Hamma, A.}
\newblock \bibinfo{title}{Stabilizer r{\'e}nyi entropy}.
\newblock \emph{\bibinfo{journal}{Physical Review Letters}} \textbf{\bibinfo{volume}{128}}, \bibinfo{pages}{050402} (\bibinfo{year}{2022}).
\newblock \eprint{2106.12587}.

\bibitem{HaugPiroli2023StabilizerEntropies}
\bibinfo{author}{Haug, T.} \& \bibinfo{author}{Piroli, L.}
\newblock \bibinfo{title}{Stabilizer entropies and nonstabilizerness monotones}.
\newblock \emph{\bibinfo{journal}{Quantum}} \textbf{\bibinfo{volume}{7}}, \bibinfo{pages}{1092} (\bibinfo{year}{2023}).
\newblock \eprint{2303.10152}.

\bibitem{LeoneBittel2024StabilizerMonotones}
\bibinfo{author}{Leone, L.} \& \bibinfo{author}{Bittel, L.}
\newblock \bibinfo{title}{Stabilizer entropies are monotones for magic-state resource theory}.
\newblock \emph{\bibinfo{journal}{Physical Review A}} \textbf{\bibinfo{volume}{110}}, \bibinfo{pages}{L040403} (\bibinfo{year}{2024}).
\newblock \eprint{2404.11652}.

\bibitem{dummit2004abstract}
\bibinfo{author}{Dummit, D.~S.}, \bibinfo{author}{Foote, R.~M.} \emph{et~al.}
\newblock \emph{\bibinfo{title}{Abstract algebra}}, vol.~\bibinfo{volume}{3} (\bibinfo{publisher}{Wiley Hoboken}, \bibinfo{year}{2004}).

\bibitem{reck1994experimental}
\bibinfo{author}{Reck, M.}, \bibinfo{author}{Zeilinger, A.}, \bibinfo{author}{Bernstein, H.~J.} \& \bibinfo{author}{Bertani, P.}
\newblock \bibinfo{title}{Experimental realization of any discrete unitary operator}.
\newblock \emph{\bibinfo{journal}{Physical review letters}} \textbf{\bibinfo{volume}{73}}, \bibinfo{pages}{58} (\bibinfo{year}{1994}).

\bibitem{Clements:16}
\bibinfo{author}{Clements, W.~R.}, \bibinfo{author}{Humphreys, P.~C.}, \bibinfo{author}{Metcalf, B.~J.}, \bibinfo{author}{Kolthammer, W.~S.} \& \bibinfo{author}{Walmsley, I.~A.}
\newblock \bibinfo{title}{Optimal design for universal multiport interferometers}.
\newblock \emph{\bibinfo{journal}{Optica}} \textbf{\bibinfo{volume}{3}}, \bibinfo{pages}{1460--1465} (\bibinfo{year}{2016}).

\bibitem{cleve1998quantum}
\bibinfo{author}{Cleve, R.}, \bibinfo{author}{Ekert, A.}, \bibinfo{author}{Macchiavello, C.} \& \bibinfo{author}{Mosca, M.}
\newblock \bibinfo{title}{Quantum algorithms revisited}.
\newblock \emph{\bibinfo{journal}{Proceedings of the Royal Society of London. Series A: Mathematical, Physical and Engineering Sciences}} \textbf{\bibinfo{volume}{454}}, \bibinfo{pages}{339--354} (\bibinfo{year}{1998}).

\bibitem{MattPeterJJiannisToAppear2}
\bibinfo{author}{Buican, M.}, \bibinfo{author}{Huston, P.} \& \bibinfo{author}{Pachos, J.~K.}
\newblock \bibinfo{title}{Reality and complexity of {$F$}-symbols in {2+1d} topological phases} (\bibinfo{year}{2026}).
\newblock \bibinfo{note}{To appear}.

\end{thebibliography}


\begin{thebibliography}{1}
\expandafter\ifx\csname url\endcsname\relax
  \def\url#1{\texttt{#1}}\fi
\expandafter\ifx\csname urlprefix\endcsname\relax\def\urlprefix{URL }\fi
\providecommand{\bibinfo}[2]{#2}
\providecommand{\eprint}[2][]{\url{#2}}

\bibitem{Byles2024}
\bibinfo{author}{Byles, L.}, \bibinfo{author}{Forbes, E.} \& \bibinfo{author}{Pachos, J.~K.}
\newblock \bibinfo{title}{Demonstration of magic state power of $d(s_3)$ anyons with two qudits}.
\newblock \emph{\bibinfo{journal}{New Journal of Physics}} \textbf{\bibinfo{volume}{28}}, \bibinfo{pages}{044501} (\bibinfo{year}{2026}).

\bibitem{dummit2004abstract}
\bibinfo{author}{Dummit, D.~S.}, \bibinfo{author}{Foote, R.~M.} \emph{et~al.}
\newblock \emph{\bibinfo{title}{Abstract algebra}}, vol.~\bibinfo{volume}{3} (\bibinfo{publisher}{Wiley Hoboken}, \bibinfo{year}{2004}).

\bibitem{BravyiKimKlieschKoenig2022Adaptive}
\bibinfo{author}{Bravyi, S.}, \bibinfo{author}{Kim, I.}, \bibinfo{author}{Kliesch, A.} \& \bibinfo{author}{Koenig, R.}
\newblock \bibinfo{title}{Adaptive constant-depth circuits for manipulating non-abelian anyons} (\bibinfo{year}{2022}).
\newblock \eprint{2205.01933}.

\bibitem{TantivasadakarnVishwanathVerresen2023Hierarchy}
\bibinfo{author}{Tantivasadakarn, N.}, \bibinfo{author}{Vishwanath, A.} \& \bibinfo{author}{Verresen, R.}
\newblock \bibinfo{title}{Hierarchy of topological order from finite-depth unitaries, measurement, and feedforward}.
\newblock \emph{\bibinfo{journal}{PRX Quantum}} \textbf{\bibinfo{volume}{4}}, \bibinfo{pages}{020339} (\bibinfo{year}{2023}).
\newblock \eprint{2209.06202}.

\bibitem{VerresenTantivasadakarnVishwanath2021Efficiently}
\bibinfo{author}{Verresen, R.}, \bibinfo{author}{Tantivasadakarn, N.} \& \bibinfo{author}{Vishwanath, A.}
\newblock \bibinfo{title}{Efficiently preparing schr{\"o}dinger's cat, fractons and non-abelian topological order in quantum devices} (\bibinfo{year}{2021}).
\newblock \eprint{2112.03061}.

\bibitem{Goel2024}
\bibinfo{author}{Goel, S.} \emph{et~al.}
\newblock \bibinfo{title}{Unveiling the non-abelian statistics of $d(s_3)$ anyons using a classical photonic simulator}.
\newblock \emph{\bibinfo{journal}{Physical Review Letters}} \textbf{\bibinfo{volume}{132}}, \bibinfo{pages}{110601} (\bibinfo{year}{2024}).

\bibitem{reck1994experimental}
\bibinfo{author}{Reck, M.}, \bibinfo{author}{Zeilinger, A.}, \bibinfo{author}{Bernstein, H.~J.} \& \bibinfo{author}{Bertani, P.}
\newblock \bibinfo{title}{Experimental realization of any discrete unitary operator}.
\newblock \emph{\bibinfo{journal}{Physical review letters}} \textbf{\bibinfo{volume}{73}}, \bibinfo{pages}{58} (\bibinfo{year}{1994}).

\end{thebibliography}


\begin{thebibliography}{0}%
\makeatletter
\providecommand \@ifxundefined [1]{%
 \@ifx{#1\undefined}
}%
\providecommand \@ifnum [1]{%
 \ifnum #1\expandafter \@firstoftwo
 \else \expandafter \@secondoftwo
 \fi
}%
\providecommand \@ifx [1]{%
 \ifx #1\expandafter \@firstoftwo
 \else \expandafter \@secondoftwo
 \fi
}%
\providecommand \natexlab [1]{#1}%
\providecommand \enquote  [1]{``#1''}%
\providecommand \bibnamefont  [1]{#1}%
\providecommand \bibfnamefont [1]{#1}%
\providecommand \citenamefont [1]{#1}%
\providecommand \href@noop [0]{\@secondoftwo}%
\providecommand \href [0]{\begingroup \@sanitize@url \@href}%
\providecommand \@href[1]{\@@startlink{#1}\@@href}%
\providecommand \@@href[1]{\endgroup#1\@@endlink}%
\providecommand \@sanitize@url [0]{\catcode `\\12\catcode `\$12\catcode `\&12\catcode `\#12\catcode `\^12\catcode `\_12\catcode `\%12\relax}%
\providecommand \@@startlink[1]{}%
\providecommand \@@endlink[0]{}%
\providecommand \url  [0]{\begingroup\@sanitize@url \@url }%
\providecommand \@url [1]{\endgroup\@href {#1}{\urlprefix }}%
\providecommand \urlprefix  [0]{URL }%
\providecommand \Eprint [0]{\href }%
\providecommand \doibase [0]{https://doi.org/}%
\providecommand \selectlanguage [0]{\@gobble}%
\providecommand \bibinfo  [0]{\@secondoftwo}%
\providecommand \bibfield  [0]{\@secondoftwo}%
\providecommand \translation [1]{[#1]}%
\providecommand \BibitemOpen [0]{}%
\providecommand \bibitemStop [0]{}%
\providecommand \bibitemNoStop [0]{.\EOS\space}%
\providecommand \EOS [0]{\spacefactor3000\relax}%
\providecommand \BibitemShut  [1]{\csname bibitem#1\endcsname}%
\let\auto@bib@innerbib\@empty
%</preamble>
\end{thebibliography}%
\putbib

\end{bibunit}

\clearpage

\appendix

\begin{bibunit}

\section*{Appendix}
\addcontentsline{toc}{section}{Appendix}

\section{Dense reduction from a four-qudit plaquette to two qutrits}
\label{sec:denseAPP}

The experimental protocol is enabled by a sequence of exact reductions of the
observables used to reconstruct the $D(S_3)$ braiding and fusion data. We begin
with the minimal four-qudit plaquette, reduce the relevant expectation values
to operators acting on two qudits, and subsequently restrict these operators
to a two-qutrit subspace. Finally, we show that the required overlaps can be
evaluated on a simple product state rather than on the original entangled
plaquette state. Each reduction preserves the squared braiding phases and
fusion amplitudes extracted by the protocol, while substantially lowering the
resources required for implementation on quantum hardware.

\subsection{From four qudits to two qudits}
\label{app:four-to-two-qudits}

We first show that the four-qudit expectation values used to reconstruct the
braiding and fusion data can be reduced exactly to expectation values on two
qudits. As described in Sec.~\ref{sec:dense}, the squared braiding phases of
the non-Abelian $G$ anyons are obtained from
\begin{equation}
    (R^i)^2
    =
    N_i
    \bra{\eta}
    F^{G}_{\rho_2}F^{G}_{\rho_1}
    A^{i}(v)
    F^{G}_{\rho_2}F^{G}_{\rho_1}
    \ket{\eta},
    \label{eq:app-R2-four-qudits}
\end{equation}
with $i\in\{+,G\}$, while the squared fusion amplitudes
$|F_{ij}|^2$ are obtained from two sets of measured expectation values
\begin{equation}
    \begin{split}
        \bra{\eta}O_{m_{ji}}\ket{\eta}= \qquad\qquad\hspace{3.3cm} \\
        \bra{\eta}F^{G}_{\rho_{2}}F^{G}_{\rho_{2}}A^{j}(v)F^{G}_{\rho_{1}}A^{G}(v)F^{G}_{\rho_{2}}A^{i}(v)F^{G}_{\rho_{2}}F^{G}_{\rho_{1}}\ket{\eta},
    \end{split}
    \label{eq:appF21}
\end{equation}
and
\begin{equation}
    \begin{split}
        \bra{\eta}O_{n_{j}}\ket{\eta}= \qquad\qquad\hspace{3.5cm} \\
        \bra{\eta}F^{G}_{\rho_{2}}F^{G}_{\rho_{2}}A^{j}(v)F^{G}_{\rho_{1}}A^{G}(v)F^{G}_{\rho_{1}}A^{j}(v)F^{G}_{\rho_{2}}F^{G}_{\rho_{2}}\ket{\eta},
    \end{split}
    \label{eq:appF22}
\end{equation}
where $i,j\in\{+,G\}$ and $N_i$ and $M_{ij}$ are the corresponding
normalisation factors. Here $\ket{\eta}$ denotes the vacuum ground state of
the minimal four-qudit $D(S_3)$ plaquette shown in
Fig.~\ref{fig:Fig1}(a), with each qudit having local dimension $d=6$.

The operator products in
Eqs.~\eqref{eq:app-R2-four-qudits}, \eqref{eq:appF21} and \eqref{eq:appF22} act non-trivially only on qudits 3 and 4 of
the plaquette~\cite{Byles2024}. Consequently, any such product can be written
as
\begin{equation}
    O=\mathbf{1}_{36}\otimes o,
\end{equation}
where $o$ acts on the Hilbert space of the remaining two qudits. Its
expectation value in the four-qudit ground state then reduces to
\begin{equation}
    \bra{\eta}O\ket{\eta}
    =
    6\sum_{g\in S_3}
    \bra{\psi_g}o\ket{\psi_g},
    \label{eq:app-reduction-to-two-qudits}
\end{equation}
where
\begin{equation}
    \ket{\psi_g}
    =
    \frac{1}{\sqrt{216}}
    \sum_{\substack{g_1,g_2\in S_3\\ g_1g_2=g}}
    \ket{g_1,g_2}.
    \label{eq:app-psi-g}
\end{equation}
Note that, with this convention, the states $\ket{\psi_g}$ are not individually
normalised. Their normalisation is incorporated into the prefactor in
Eq.~\eqref{eq:app-reduction-to-two-qudits}. Thus, all four-qudit plaquette
expectation values required by the protocol can be evaluated using operators
acting on only two $d=6$ qudits.

\subsection{From two qudits to two qutrits}
\label{app:two-qudits-to-qutrits}

We next exploit the semidirect-product structure
$S_3\simeq \mathbb{Z}_3\rtimes\mathbb{Z}_2$ to reduce each local
six-dimensional qudit to a qutrit~\cite{dummit2004abstract}. Ordering the
group-element basis as
\begin{equation}
    \{e,c,c^2\}\oplus\{t,tc,tc^2\},
\end{equation}
separates the $\mathbb{Z}_3$ subgroup from its complementary coset. This decomposition also underlies efficient constructions of $D(S_3)$ states on quantum hardware~\cite{BravyiKimKlieschKoenig2022Adaptive, TantivasadakarnVishwanathVerresen2023Hierarchy, VerresenTantivasadakarnVishwanath2021Efficiently}.

For the braiding and fusion processes considered here, the relevant operators
in the $\{+,G\}$ sector have support only on the
$\mathbb{Z}_3=\{e,c,c^2\}$ block. Here
\begin{equation}
    \ket{+}\equiv
    \frac{\ket A+\ket B}{\sqrt{2}}
\end{equation}
denotes the normalised Abelian-sector superposition accessed by the reduced
protocol. Each term in the corresponding two-qudit operator products can be
written in the form
\begin{equation}
    (\mathcal{B}\oplus\mathbf{0}_3)
    \otimes
    (\mathcal{C}\oplus\mathbf{0}_3),
    \label{eq:app-op-product}
\end{equation}
where $\mathcal{B}$ and $\mathcal{C}$ act on the $\mathbb{Z}_3$ subspace.
The complementary $\{t,tc,tc^2\}$ block therefore does not contribute to the
group-averaged expectation values. In particular,
\begin{equation}
\begin{split}
    &\sum_{g\in S_3}
    \bra{\psi_g}
    (\mathcal{B}\oplus\mathbf{0}_3)
    \otimes
    (\mathcal{C}\oplus\mathbf{0}_3)
    \ket{\psi_g}
    \\
    &\hspace{2cm}
    =
    \frac{1}{72}
    \sum_{g\in\mathbb{Z}_3}
    \bra{\Psi_g}
    \mathcal{B}\otimes\mathcal{C}
    \ket{\Psi_g},
\end{split}
\label{eq:app-qudit-to-qutrit-reduction}
\end{equation}
where the normalised two-qutrit states
\begin{equation}
    \ket{\Psi_g}
    =
    \frac{1}{\sqrt{3}}
    \sum_{\substack{g_1,g_2\in\mathbb{Z}_3\\g_1g_2=g}}
    \ket{g_1,g_2}
    \label{eq:app-Psi-g}
\end{equation}
have definite total $\mathbb{Z}_3$ flux $g$. The factor $1/72$ follows from
the normalisation of the two-qudit states $\ket{\psi_g}$ in
Eq.~\eqref{eq:app-psi-g}.

Consequently, all expectation values entering
Eqs.~\eqref{eq:app-R2-four-qudits}, \eqref{eq:appF21} and \eqref{eq:appF22} can be evaluated using the reduced two-qutrit
operators
\begin{align}
    \mathcal{F}^{G}_{\rho_1}
    &=
    Z\otimes X+Z^2\otimes X^2,
    \label{eq:app-reduced-ribbon-1}\\
    \mathcal{F}^{G}_{\rho_2}
    &=
    X\otimes Z^2+X^2\otimes Z,
    \label{eq:app-reduced-ribbon-2}\\
    \mathcal{A}^{+}(v)
    &=
    \frac{1}{3}
    \left(
        \mathbf{1}_9
        +X\otimes X^2
        +X^2\otimes X
    \right),
    \label{eq:app-reduced-Aplus}\\
    \mathcal{A}^{G}(v)
    &=
    \frac{1}{3}
    \left(
        2\mathbf{1}_9
        -X\otimes X^2
        -X^2\otimes X
    \right).
    \label{eq:app-reduced-AG}
\end{align}
Here
\begin{equation}
    X=
    \begin{pmatrix}
        0&0&1\\
        1&0&0\\
        0&1&0
    \end{pmatrix},
    \qquad
    Z=
    \begin{pmatrix}
        1&0&0\\
        0&\omega&0\\
        0&0&\bar{\omega}
    \end{pmatrix},
    \qquad
    \omega=e^{2\pi i/3},
\end{equation}
are the qutrit shift and clock operators, satisfying
$X^3=Z^3=\mathbf{1}_3$ and $ZX=\omega XZ$.

The operators $\mathcal{F}^{G}_{\rho_1}$ and
$\mathcal{F}^{G}_{\rho_2}$ retain the action of the two intersecting
$G$-anyon ribbons, while $\mathcal{A}^{+}(v)$ and
$\mathcal{A}^{G}(v)$ resolve the corresponding Abelian and non-Abelian charge
sectors. Thus, the four-qudit observables required to reconstruct the
braiding and fusion data are reduced exactly to operators on two qutrits. In
Appendix ~\ref{sec:RFqutrit}, we verify explicitly that these reduced operators
retain the fusion rules and braiding phases of the relevant
$\{A,B,G\}$ subcategory of $D(S_3)$.

\subsection{From entangled to product states}
\label{sec:densePROD}

The final reduction replaces the ensemble of entangled two-qutrit states
$\ket{\Psi_g}$ defined in Eq.~\eqref{eq:app-Psi-g} by a single, experimentally
accessible product state. The states $\ket{\Psi_g}$ have definite total
$\mathbb{Z}_3$ flux $g$, and therefore their expectation values depend only
on matrix elements that preserve this flux.

To make this structure explicit, write a general two-qutrit operator
$\mathcal{O}$ as
\begin{align}
    \mathcal{O}
    &=
    \sum_{\substack{
        g_1g_2=g_3g_4\\
        g_1,\ldots,g_4\in\mathbb{Z}_3
    }}
    p_{g_1g_2g_3g_4}
    \ket{g_1,g_2}\bra{g_3,g_4}
    \nonumber\\
    &\quad+
    \sum_{\substack{
        g_1g_2\neq g_3g_4\\
        g_1,\ldots,g_4\in\mathbb{Z}_3
    }}
    q_{g_1g_2g_3g_4}
    \ket{g_1,g_2}\bra{g_3,g_4},
    \label{eq:Onm}
\end{align}
where the coefficients $p_{g_1g_2g_3g_4}$ and
$q_{g_1g_2g_3g_4}$ multiply the flux-preserving and flux-changing matrix
elements, respectively. Since $\ket{\Psi_g}$ has fixed total flux, matrix
elements connecting different flux sectors do not contribute, and hence
\begin{equation}
    \sum_{g\in\mathbb{Z}_3}
    \bra{\Psi_g}\mathcal{O}\ket{\Psi_g}
    =
    \frac{1}{3}
    \sum_{\substack{
        g_1g_2=g_3g_4\\
        g_1,\ldots,g_4\in\mathbb{Z}_3
    }}
    p_{g_1g_2g_3g_4}.
    \label{eq:flux-conserving-average}
\end{equation}

Consider the normalised equal-superposition state
\begin{equation}
    \ket{\xi}
    =
    \frac{1}{3}
    \sum_{g_1,g_2\in\mathbb{Z}_3}
    \ket{g_1,g_2}.
    \label{eq:xiAPP}
\end{equation}
This state factorizes as
\begin{equation}
    \ket{\xi}
    =
    \left(
        \frac{1}{\sqrt{3}}
        \sum_{g_1\in\mathbb{Z}_3}\ket{g_1}
    \right)
    \otimes
    \left(
        \frac{1}{\sqrt{3}}
        \sum_{g_2\in\mathbb{Z}_3}\ket{g_2}
    \right),
\end{equation}
and is therefore a product state of the two qutrits. Its expectation value
contains both the flux-preserving and flux-changing components of
$\mathcal O$:
\begin{align}
    9\bra{\xi}\mathcal{O}\ket{\xi}
    &=
    \sum_{\substack{
        g_1g_2=g_3g_4\\
        g_1,\ldots,g_4\in\mathbb{Z}_3
    }}
    p_{g_1g_2g_3g_4}
    \nonumber\\
    &\quad+
    \sum_{\substack{
        g_1g_2\neq g_3g_4\\
        g_1,\ldots,g_4\in\mathbb{Z}_3
    }}
    q_{g_1g_2g_3g_4}.
    \label{eq:xi-general-expectation}
\end{align}

For the particular operator products entering the braiding and fusion
overlaps in Eqs.~\eqref{eq:app-R2-four-qudits}, \eqref{eq:appF21} and \eqref{eq:appF22}, direct evaluation shows that the total
flux-changing contribution cancels
\begin{equation}
    \sum_{\substack{
        g_1g_2\neq g_3g_4\\
        g_1,\ldots,g_4\in\mathbb{Z}_3
    }}
    q_{g_1g_2g_3g_4}
    =
    0.
    \label{eq:flux-changing-cancellation}
\end{equation}
Equations~\eqref{eq:flux-conserving-average}-\eqref{eq:flux-changing-cancellation}
then imply
\begin{equation}
    \sum_{g\in\mathbb{Z}_3}
    \bra{\Psi_g}\mathcal{O}\ket{\Psi_g}
    =
    3\bra{\xi}\mathcal{O}\ket{\xi}.
    \label{eq:Psi-to-xi}
\end{equation}

Combining Eq.~\eqref{eq:Psi-to-xi} with the four-to-two-qudit reduction in
Eq.~\eqref{eq:app-reduction-to-two-qudits} and the subsequent restriction to
the $\mathbb{Z}_3$ block in
Eq.~\eqref{eq:app-qudit-to-qutrit-reduction}, including the normalization
factors absorbed into the definitions of the reduced operators, gives
\begin{equation}
    \bra{\eta}O\ket{\eta}
    =
    \frac{1}{4}\bra{\xi}\mathcal{O}\ket{\xi}.
    \label{eq:final-product-state-reduction}
\end{equation}
Thus, every four-qudit plaquette overlap required to reconstruct the
braiding phases and fusion amplitudes can be evaluated using reduced
operators acting on two qutrits prepared in the simple product state
$\ket{\xi}$. This exact product-state reduction is what makes the protocol
directly implementable on the quantum hardware described in the Methods.

\subsection{Anyonic content of the reduced two-qutrit operators}
\label{sec:appqut}

The dense reduction is useful only if the resulting two-qutrit operators
retain the fusion and braiding data of the $\{A,B,G\}$ fusion subcategory of
$D(S_3)$. A detailed derivation of the corresponding ribbon algebra in the
full quantum-double construction was given in Ref.~\cite{Goel2024}. Here we
record only the identities needed to establish the consistency of the reduced
description used in the present experiment.

\subsubsection{Fusion relations}

In the ordered group basis
$\{e,c,c^2,t,tc,tc^2\}$, the ribbon operators creating the Abelian charges
$A$ and $B$ can be written as
\begin{equation}
    F^A=\mathbf{1}_6,
    \qquad
    F^B=
    \operatorname{diag}(1,1,1,-1,-1,-1).
\end{equation}
Their sum therefore has support only on the
$\mathbb{Z}_3=\{e,c,c^2\}$ block. In the reduced two-qutrit description, the
corresponding composite ribbon is
\begin{equation}
    \mathcal{F}^{+}
    \equiv
    \mathcal{F}^{A}+\mathcal{F}^{B}
    =
    2\mathbf{1}_3\otimes\mathbf{1}_3,
    \label{eq:app-Fplus}
\end{equation}
where $+$ denotes the unnormalised composite charge $A+B$.

Together with the reduced $G$-ribbon operators in
Eqs.~\eqref{eq:app-reduced-ribbon-1} and
\eqref{eq:app-reduced-ribbon-2}, direct multiplication gives
\begin{align}
    \left(\mathcal{F}^{+}\right)^2
    &=
    2\mathcal{F}^{+},
    \label{eq:app-fusion-plus-plus}\\
    \mathcal{F}^{+}\mathcal{F}^{G}_{\rho_k}
    &=
    2\mathcal{F}^{G}_{\rho_k},
    \qquad k=1,2,
    \label{eq:app-fusion-plus-G}\\
    \left(\mathcal{F}^{G}_{\rho_k}\right)^2
    &=
    \mathcal{F}^{+}+\mathcal{F}^{G}_{\rho_k},
    \qquad k=1,2.
    \label{eq:app-fusion-G-G}
\end{align}
These identities reproduce
\begin{equation}
    +\times +=2+,
    \qquad
    +\times G=2G,
    \qquad
    G\times G=A+B+G,
\end{equation}
respectively
The factors of two in the first two relations arise because $+$ denotes the
unnormalised composite $A+B$. Thus, the reduced ribbon operators retain the
fusion algebra of the relevant $D(S_3)$ anyons.

\subsubsection{Braiding phases}

The braiding information is encoded in the non-commutativity of the two
intersecting $G$-ribbon operators~\cite{Goel2024}. For every non-zero matrix element in the
two-qutrit group basis, direct multiplication gives
\begin{equation}
\begin{split}
    &\bra{g_1,g_2}
    \mathcal{F}^{G}_{\rho_1}\mathcal{F}^{G}_{\rho_2}
    \ket{h_1,h_2}
    \\
    &\quad=
    \begin{cases}
        \bar{\omega}\,
        \bra{g_1,g_2}
        \mathcal{F}^{G}_{\rho_2}\mathcal{F}^{G}_{\rho_1}
        \ket{h_1,h_2},
        & g_1g_2=h_1h_2,\\[2mm]
        \omega\,
        \bra{g_1,g_2}
        \mathcal{F}^{G}_{\rho_2}\mathcal{F}^{G}_{\rho_1}
        \ket{h_1,h_2},
        & g_1g_2\neq h_1h_2,
    \end{cases}
\end{split}
\label{eq:app-reduced-braiding-relation}
\end{equation}
where $\omega=e^{2\pi i/3}$.

The condition $g_1g_2=h_1h_2$ preserves the plaquette flux and therefore
corresponds to the trivial-flux fusion channels $A$ and $B$. These channels
acquire the common squared braiding phase $\bar{\omega}$. By contrast,
$g_1g_2\neq h_1h_2$ changes the plaquette flux and corresponds to the
non-trivial-flux channel $G$, which acquires the phase $\omega$. Hence,
\begin{equation}
    (R^A)^2=(R^B)^2=\bar{\omega},
    \qquad
    (R^G)^2=\omega,
    \label{eq:app-recovered-R2}
\end{equation}
in agreement with the $D(S_3)$ braiding data in Eq.~\eqref{eq:RF}. The
reduced two-qutrit operators therefore preserve both the fusion algebra and
the squared braiding phases reconstructed experimentally.

\section{Recovery of the anyonic data}
\label{sec:RFqutrit}

The preceding section established an exact reduction of the four-qudit
plaquette overlaps to expectation values of operators acting on two qutrits.
We now show how the experimentally accessible measurements in the
$\{+,G\}$ sectors recover the braiding and fusion data in the physical
anyon basis $\{A,B,G\}$. The reduced protocol does not resolve the $A$ and
$B$ sectors independently, but instead accesses their coherent superposition
\begin{equation}
    \ket{+}
    \equiv
    \frac{\ket A+\ket B}{\sqrt{2}}.
\end{equation}
For braiding, this is sufficient because the $A$ and $B$ fusion channels have
the same squared braiding phase. For fusion, the measured $+$- and
$G$-sector overlaps, together with the symmetry and unitarity of the
$D(S_3)$ $F$-symbol, determine the squared amplitudes in the full
$\{A,B,G\}$ basis.

\subsection{Squared braiding phases from qutrit measurements}
\label{app:R-from-qutrits}

We first define the states whose overlap gives the squared braiding phase.
Let
\begin{align}
    \ket{\widetilde{\phi}_{12}(i)}
    &\equiv
    A^{i}(v)
    F^{G}_{\rho_1}F^{G}_{\rho_2}
    \ket{\eta},
    \label{eq:phi12-unnormalised}\\
    \ket{\widetilde{\phi}_{21}(i)}
    &\equiv
    A^{i}(v)
    F^{G}_{\rho_2}F^{G}_{\rho_1}
    \ket{\eta},
    \label{eq:phi21-unnormalised}
\end{align}
where $i\in\{A,B,G\}$. Here $\ket{\eta}$ is the vacuum state of the
minimal four-qudit plaquette, the two ribbon operators create the same pair
of $G$ anyons in opposite orders, and $A^{i}(v)$ resolves their total
fusion channel at the shared endpoint. We denote the corresponding
normalised states by
\begin{equation}
    \ket{\phi_{\mu}(i)}
    =
    \frac{\ket{\widetilde{\phi}_{\mu}(i)}}
    {\sqrt{
        \braket{\widetilde{\phi}_{\mu}(i)|
        \widetilde{\phi}_{\mu}(i)}
    }}.
    \label{eq:normalised-phi-states}
\end{equation}
with $\mu\in\{12,21\}$.

The two states differ by a double exchange of the $G$ anyons. With the
orientation convention used in Fig.~\ref{fig:Fig1}(b), their relation in a
fixed fusion channel is
\begin{equation}
    \ket{\phi_{21}(i)}
    =
    (R^{i})^{2}\ket{\phi_{12}(i)},
    \label{eq:phi-braiding-relation}
\end{equation}
and therefore
\begin{equation}
    (R^{i})^{2}
    =
    \braket{\phi_{12}(i)|\phi_{21}(i)}.
    \label{eq:R2-as-phi-overlap}
\end{equation}

The reduced protocol accesses the normalised Abelian-sector superpositions
\begin{equation}
    \ket{\phi_{\mu}(+)}
    \equiv
    \frac{1}{\sqrt{2}}
    \left[
        \ket{\phi_{\mu}(A)}
        +
        \ket{\phi_{\mu}(B)}
    \right].
    \label{eq:phi-plus-definition}
\end{equation}
with $\mu\in\{12,21\}$. Using Eq.~\eqref{eq:phi-braiding-relation}, the state obtained after the
double exchange is
\begin{equation}
\begin{split}
    \ket{\phi_{21}(+)}
    =
    \frac{1}{\sqrt{2}}
    \big[
        &(R^{A})^{2}\ket{\phi_{12}(A)}
        \\
        &+
        (R^{B})^{2}\ket{\phi_{12}(B)}
    \big].
\end{split}
\label{eq:phi-plus-after-braid}
\end{equation}
On the other hand, the reduced $+$ sector transforms with a single measured
phase,
\begin{equation}
\begin{split}
    \ket{\phi_{21}(+)}
    &=
    (R^{+})^{2}\ket{\phi_{12}(+)}
    \\
    &=
    \frac{(R^{+})^{2}}{\sqrt{2}}
    \left[
        \ket{\phi_{12}(A)}
        +
        \ket{\phi_{12}(B)}
    \right].
\end{split}
\label{eq:phi-plus-single-phase}
\end{equation}
Since the $A$ and $B$ fusion channels are orthogonal, comparison of
Eqs.~\eqref{eq:phi-plus-after-braid} and
\eqref{eq:phi-plus-single-phase} gives
\begin{equation}
    (R^{A})^{2}
    =
    (R^{B})^{2}
    =
    (R^{+})^{2}.
    \label{eq:RAB-from-Rplus}
\end{equation}

Using the dense reduction derived in Sec.~\ref{sec:denseAPP}, the common
Abelian-sector phase can be evaluated on the two-qutrit product state
$\ket{\xi}$ as
\begin{equation}
\begin{split}
    (R^{+})^{2}
    =
    \mathcal{N}_{+}
    \bra{\xi}
    \mathcal{F}^{G}_{\rho_2}
    \mathcal{F}^{G}_{\rho_1}
    \mathcal{A}^{+}(v)
    \mathcal{F}^{G}_{\rho_2}
    \mathcal{F}^{G}_{\rho_1}
    \ket{\xi},
\end{split}
\label{eq:app-Rplus-qutrit-overlap}
\end{equation}
where $\mathcal{N}_{+}=1/2$ includes the normalisation associated with the composite
$+$ projection and the scaled implementation of the ribbon operators. This
is the non-unitary expectation value reconstructed by the adapted Hadamard
test described in Sec.~\ref{sec:braiding-extraction}.

The $G$-sector phase is obtained from the separately measured
identity-channel overlap rather than from an additional controlled
$G$-projection circuit. With the normalisation conventions of the protocol,
\begin{equation}
    (R^{G})^{2}
    =
    (R^{\mathbf{1}})^{2}
    -
    (R^{+})^{2},
    \label{eq:app-RG-identity-subtraction}
\end{equation}
in agreement with Eq.~\eqref{eq:RG-measurement}. Thus, the two reduced
measurements $(R^{+})^{2}$ and $(R^{\mathbf{1}})^{2}$ recover all three
squared braiding phases in the physical $\{A,B,G\}$ basis.

\subsection{Squared fusion amplitudes from qutrit measurements}
\label{sec:Fqutrit}

We now show how the experimentally accessible two-qutrit overlaps determine
the squared fusion amplitudes in the physical anyon basis
$\{A,B,G\}$. We write
\begin{equation}
    F\equiv F^{G}_{GGG},
    \qquad
    \Phi_{ij}\equiv |F_{ij}|^2,
    \label{eq:app-Phi-definition}
\end{equation}
with $i,j\in\{A,B,G\}$. The four-qudit plaquette construction relating overlaps between alternative
fusion histories to the corresponding $F$-symbols was derived in
Ref.~\cite{Byles2024}. We denote the unnormalised overlap associated with
intermediate fusion channels $i$ and $j$ by
\begin{equation}
    \widetilde m_{ji}
    =
    \bra{\eta}O_{m_{ji}}\ket{\eta},
    \label{eq:app-mtilde-definition}
\end{equation}
with $i,j\in\{A,B,G\}$, where
\begin{equation}
    O_{m_{ji}}
    =
    F^{G}_{\rho_{2}}
    F^{G}_{\rho_{2}}
    A^{j}(v)
    F^{G}_{\rho_{1}}
    A^{G}(v)
    F^{G}_{\rho_{2}}
    A^{i}(v)
    F^{G}_{\rho_{2}}
    F^{G}_{\rho_{1}} .
    \label{eq:appmji}
\end{equation}
The corresponding normalisation of the $j$-resolved fusion history is
\begin{equation}
    \widetilde n_{j}
    =
    \bra{\eta}O_{n_{j}}\ket{\eta},
    \label{eq:app-ntilde-definition}
\end{equation}
with $j\in\{A,B,G\}$ and
\begin{equation}
    O_{n_{j}}
    =
    F^{G}_{\rho_{2}}
    F^{G}_{\rho_{2}}
    A^{j}(v)
    F^{G}_{\rho_{1}}
    A^{G}(v)
    F^{G}_{\rho_{1}}
    A^{j}(v)
    F^{G}_{\rho_{2}}
    F^{G}_{\rho_{2}} .
    \label{eq:appnj}
\end{equation}
The resulting overlap ratio is related to the fusion matrix element by
\begin{equation}
    (F_{ij})^2
    =
    \left(R^{Gj}_{G}\right)^*
    \frac{\widetilde m_{ji}}{\widetilde n_j}.
    \label{eq:app-F-overlap-relation}
\end{equation}
Here
\begin{equation}
    R^{Gj}_{G}
    =
    \begin{cases}
        1, & j=A,B,\\
        \omega, & j=G,
    \end{cases}
    \qquad
    \omega=e^{2\pi i/3},
    \label{eq:app-fusion-braid-factor}
\end{equation}
is the known phase generated by the crossing of the ribbon histories.
Removing this phase and taking the modulus gives the squared fusion
amplitude $\Phi_{ij}=|F_{ij}|^2$.

We next apply the dense reduction of Sec.~\ref{sec:denseAPP}. Introducing
the combined charge projection
$A^{+}(v)=A^{A}(v)+A^{B}(v)$, the experimentally accessible sectors are
$i,j\in\{+,G\}$. For the overlap quantities, the four-qudit expectation
values reduce to
\begin{align}
    \widetilde m_{ji}
    =
    \bra{\eta}O_{m_{ji}}\ket{\eta}
%    \nonumber\\
    =
    \frac{1}{4}
    \bra{\xi}\mathcal{O}_{m_{ji}}\ket{\xi}
    \equiv
    \frac{1}{4}m_{ji},
    \label{eq:app-m-reduction}
\end{align}
with $i,j\in\{+,G\}$, where $\mathcal{O}_{m_{ji}}$ denotes the corresponding product of reduced
two-qutrit ribbon and projection operators. Similarly,
\begin{align}
    \widetilde n_j
    =
    \bra{\eta}O_{n_j}\ket{\eta}
%    \nonumber\\
    =
    \frac{1}{4}
    \bra{\xi}\mathcal{O}_{n_j}\ket{\xi}
    \equiv
    \frac{1}{4}n_j,
    \label{eq:app-n-reduction}
\end{align}
with $j\in\{+,G\}$.
The reduced quantities
$m_{++}$, $m_{+G}$, $m_{G+}$, $m_{GG}$, $n_{+}$ and $n_{G}$ can
therefore be evaluated using the two-qutrit product state $\ket{\xi}$.
The common factor of $1/4$ cancels from every normalised overlap,
\begin{equation}
    \frac{\widetilde m_{ji}}{\widetilde n_j}
    =
    \frac{m_{ji}}{n_j}.
    \label{eq:app-ratio-reduction}
\end{equation}
It is therefore sufficient in what follows to work with the reduced
two-qutrit quantities $m_{ji}$ and $n_j$. 

We parameterise the $F$-symbol in the real symmetric gauge used throughout
this work as
\begin{equation}
    F
    =
    \begin{pmatrix}
        f_{1} & f_{2} & f_{3}\\
        f_{2} & f_{4} & f_{5}\\
        f_{3} & f_{5} & f_{6}
    \end{pmatrix},
    \label{eq:app-F-parametrisation}
\end{equation}
where the rows and columns are ordered as $A,B,G$. The corresponding matrix
of elementwise squared magnitudes is
\begin{equation}
    \Phi
    =
    \begin{pmatrix}
        \phi_{1} & \phi_{2} & \phi_{3}\\
        \phi_{2} & \phi_{4} & \phi_{5}\\
        \phi_{3} & \phi_{5} & \phi_{6}
    \end{pmatrix},
    \qquad
    \phi_k\equiv |f_k|^2.
    \label{eq:app-Phi-parametrisation}
\end{equation}
Since the chosen gauge is real, $\phi_k=f_k^2$, although the
squared-magnitude notation makes explicit the quantities reconstructed by
the protocol.

Expanding the experimentally accessible $+$- and $G$-sector overlaps using
Eq.~\eqref{eq:app-F-overlap-relation} gives
\begin{equation}
    \phi_6
    =
    \frac{m_{GG}}{\omega n_G},
    \label{eq:app-phi6}
\end{equation}
together with
\begin{align}
    m_{++}
    &=
    n_A(\phi_1+\phi_2)
    +
    n_B(\phi_2+\phi_4),
    \label{eq:m_pp}\\
    m_{+G}
    &=
    n_A\phi_3+n_B\phi_5,
    \label{eq:m_pg}\\
    m_{G+}
    &=
    \omega n_G(\phi_3+\phi_5).
    \label{eq:m_gp}
\end{align}

To expose the information contained in the unresolved Abelian sector, we
complete the experimentally accessible $\{+,G\}$ sectors by introducing the
orthonormal combinations
\begin{equation}
    \ket{+}
    =
    \frac{\ket A+\ket B}{\sqrt{2}},
    \qquad
    \ket{-}
    =
    \frac{\ket A-\ket B}{\sqrt{2}}.
    \label{eq:app-plus-minus-basis}
\end{equation}
Correspondingly, we define
\begin{equation}
    A^{\pm}(v)
    =
    A^{A}(v)\pm A^{B}(v).
    \label{eq:app-plus-minus-operators}
\end{equation}
Here $A^{+}(v)$ is the combined Abelian-sector operation used in the
experiment, whereas $A^{-}(v)$ is an auxiliary signed combination introduced
only to analyse the unresolved $A$ and $B$ channels.

To determine the mixed $+$ and $-$ overlaps, define
\begin{align}
    O_{1,i}
    &=
    A^{G}(v)F^{G}_{\rho_2}A^{i}(v)
    F^{G}_{\rho_2}F^{G}_{\rho_1},
    \label{eq:app-O1}\\
    O_{2,j}
    &=
    A^{G}(v)F^{G}_{\rho_1}A^{j}(v)
    F^{G}_{\rho_2}F^{G}_{\rho_2},
    \label{eq:app-O2}
\end{align}
so that
\begin{equation}
    O_{m_{ji}}=O_{2,j}^{\dagger}O_{1,i}.
\end{equation}
Under the decomposition
$\mathcal H_{S_3}=\mathcal H_{\mathbb Z_3}\oplus
\mathcal H_{t\mathbb Z_3}$, direct expansion shows that
$O_{1,-}$ and $O_{2,+}$ have support on orthogonal blocks, as do
$O_{1,+}$ and $O_{2,-}$. Consequently,
\begin{equation}
    m_{+-}=m_{-+}=0.
    \label{eq:app-mixed-overlaps-zero}
\end{equation}
The corresponding mixed normalisation overlap also vanishes,
\begin{align}
    n_{+-} &= 4\bra{\eta}O_{2,+}^{\dagger}O_{2,-}\ket{\eta}, \\
    &=n_A-n_B
    =
    0,
\end{align}
and hence
\begin{equation}
    n_A=n_B=\frac{n_+}{2}.
    \label{eq:app-nAB}
\end{equation}

In terms of the entries of $\Phi$, the two mixed overlaps are
\begin{align}
    m_{+-}
    &=
    n_A(\phi_1-\phi_2)
    +
    n_B(\phi_2-\phi_4)
    =
    0,
    \label{eq:m_pm}\\
    m_{-+}
    &=
    n_A(\phi_1+\phi_2)
    -
    n_B(\phi_2+\phi_4)
    =
    0.
    \label{eq:m_mp}
\end{align}
Using Eq.~\eqref{eq:app-nAB}, either relation gives
\begin{equation}
    \phi_1=\phi_4.
    \label{eq:app-phi14}
\end{equation}
The additional equality $\phi_2=\phi_1$ is the repeated $A$--$B$ block of
the exact $D(S_3)$ $F$-symbol in the chosen gauge. Using this model-specific
property, the Abelian block therefore satisfies
\begin{equation}
    \phi_1=\phi_2=\phi_4.
    \label{eq:phi124}
\end{equation}
Equation~\eqref{eq:m_pp} then reduces to
\begin{equation}
    m_{++}=2n_+\phi_1,
\end{equation}
and hence
\begin{equation}
    \phi_1
    =
    \frac{m_{++}}{2n_+}.
    \label{eq:app-phi1}
\end{equation}

Unitarity of $F$ implies that the squared magnitudes in each row sum to
unity. The first two rows of Eq.~\eqref{eq:app-Phi-parametrisation}
therefore satisfy
\begin{equation}
    \phi_1+\phi_2+\phi_3
    =
    \phi_2+\phi_4+\phi_5
    =
    1.
\end{equation}
Together with Eq.~\eqref{eq:phi124}, this gives
\begin{equation}
    \phi_3=\phi_5.
    \label{eq:app-phi35}
\end{equation}
The two equations \eqref{eq:m_pg} and \eqref{eq:m_gp} provide independent estimates of the squared amplitudes,
\begin{equation}
    \phi_{3}^{(+G)}
    =
    \frac{m_{+G}}{n_+},
    \qquad
    \phi_{3}^{(G+)}
    =
    \frac{m_{G+}}{2\omega n_G}.
    \label{eq:app-phi35-estimates}
\end{equation}
% Using the relations
% $\phi_1=\phi_2=\phi_4$, the corresponding unconstrained
% squared-amplitude matrix is
% \begin{equation}
%     \Phi_{\mathrm{unc}}
%     =
%     \begin{pmatrix}
%         \phi_1 & \phi_1 & \phi_{3}\\
%         \phi_1 & \phi_1 & \phi_{5}\\
%         \phi_{3} & \phi_{} & \phi_6
%     \end{pmatrix}.
%     \label{eq:app-Phi-unconstrained}
% \end{equation}
% In the real symmetric gauge of the exact $D(S_3)$ $F$-symbol,
% $F=F^{T}$ and therefore $\Phi=\Phi^{T}$. Accordingly, the two
% theoretical expressions coincide,
% $\phi_{3}^{(+G)}=\phi_{3}^{(G+)}$. 
As they are obtained from distinct circuit realisations and may differ because of finite sampling we impose the symmetrisation
\begin{equation}
    \phi_3
    =
    \frac{1}{2}
    \left[
        \phi_{3}^{(+G)}
        +
        \phi_{3}^{(G+)}
    \right]
    =
    \frac{1}{2}
    \left(
        \frac{m_{+G}}{n_+}
        +
        \frac{m_{G+}}{2\omega n_G}
    \right).
    \label{eq:app-phi35-symmetrised}
\end{equation}

% Experimentally, the two overlap ratios are reconstructed from the
% corresponding post-selected probabilities. The symmetrised estimator is
% therefore
% \begin{equation}
%     {\phi}_3
%     =
%     {\phi}_5
%     =
%     \frac{1}{2}
%     \left[
%         \sqrt{\frac{p(m_{+G})}{p(n_+)}}
%         +
%         \frac{1}{2}
%         \sqrt{\frac{p(m_{G+})}{p(n_G)}}
%     \right].
%     \label{eq:app-phi35-probabilities}
% \end{equation}

The squared fusion-amplitude matrix can therefore be written in terms of
three independent quantities as
\begin{equation}
    \Phi
    =
    \begin{pmatrix}
        a & a & b\\
        a & a & b\\
        b & b & c
    \end{pmatrix},
    \label{eq:app-Phi-abc}
\end{equation}
with $a=\phi_1$, $b=\phi_3$,
$c=\phi_6$, where
\begin{equation}
    a
    =
    \frac{m_{++}}{2n_{+}},
    \,\,
    b
    =
    \frac{1}{2}
    \left(
        \frac{m_{+G}}{n_{+}}
        +
        \frac{m_{G+}}{2\omega n_{G}}
    \right),
    \,\,
    c
    =
    \frac{m_{GG}}{\omega n_{G}}.
    \label{eq:app-F-amplitude-ratios}
\end{equation}
Thus, the six reduced quantities
$m_{++}$, $m_{+G}$, $m_{G+}$, $m_{GG}$, $n_{+}$ and $n_G$
determine the three independent squared amplitudes $a$, $b$ and $c$.

The matrix $\Phi$ determines only the magnitudes of the entries of $F$.
Since the fusion spaces $V^{i}_{GG}$ are one-dimensional for
$i\in\{A,B,G\}$, a change of vertex gauge amounts to rephasing each
trivalent vertex $G\times G\rightarrow i$. Restricting to gauges in which
$F$ remains real, these rephasings reduce to independent signs
$s_i=\pm1$. Introducing
\begin{equation}
    D_{\boldsymbol{s}}
    =
    \operatorname{diag}(s_A,s_B,s_G),
\end{equation}
the real symmetric representatives compatible with a reference matrix
$F_0$ can be written as
\begin{equation}
    F_{\boldsymbol{s},\epsilon}
    =
    \epsilon
    D_{\boldsymbol{s}}F_0D_{\boldsymbol{s}},
    \qquad
    \epsilon=\pm1.
    \label{eq:app-F-gauge-signs}
\end{equation}
The diagonal matrices implement the corresponding sign changes of the
fusion-channel basis on the two fusion trees, while $\epsilon$ represents
their relative overall sign. Since
$D_{\boldsymbol{s}}$ and $-D_{\boldsymbol{s}}$ generate the same
transformation $D_{\boldsymbol{s}}F_0D_{\boldsymbol{s}}$, the three vertex
signs produce four distinct channel-sign patterns. Including the overall
sign $\epsilon$ gives eight admissible sign assignments, grouped into four
pairs that differ only by an overall sign. These matrices are related by
fusion-vertex gauge transformations and therefore represent
gauge-equivalent real forms of the same $D(S_3)$ $F$-symbol. We apply the corresponding gauge-related sign patterns to the experimentally reconstructed
magnitudes, although finite sampling means that the resulting matrices
are not exactly unitary.

\section{Circuit design for encoded qutrit operations}
\label{app:gate-design}

We now describe the circuit decompositions used to implement the reduced
two-qutrit protocol in the three-qubit encoding introduced in
Eq.~\eqref{eq:encode}. The required operations comprise the qutrit Fourier
transform, preparation of the product state $\ket{\xi}$, and the
ancilla-assisted implementations of the charge projections and ribbon
operators. Unitary qutrit transformations are compiled into single-qubit
rotations and two-qubit Givens rotations that preserve the
unit-Hamming-weight subspace, whereas the non-unitary primitives are realised
through unitary dilations followed by heralding on ancillary outcomes.

\subsection{Qutrit Fourier transform}
\label{app:QFT}

In the logical basis $\{\ket{e},\ket{c},\ket{c^2}\}$, the qutrit Fourier
transform is
\begin{equation}
    U_{\mathrm Q}
    =
    \frac{1}{\sqrt{3}}
    \begin{pmatrix}
        1 & 1 & 1\\
        1 & \omega & \omega^2\\
        1 & \omega^2 & \omega
    \end{pmatrix},
    \qquad
    \omega=e^{2\pi i/3}.
    \label{eq:app-qutrit-qft}
\end{equation}
To implement $U_{\mathrm Q}$ in the three-qubit encoding, we decompose it
using the Reck construction~\cite{reck1994experimental}, which expresses a
general $\mathrm{U}(3)$ transformation as a sequence of $\mathrm{U}(2)$
rotations acting on pairs of logical basis states. Within the
unit-Hamming-weight encoding, each such rotation is realised by a two-qubit
gate acting on the subspace spanned by $\ket{01}$ and $\ket{10}$, as
described in Eq.~\eqref{eq:U2}.

\begin{figure}
    \centering
    \includegraphics[width=\columnwidth]
    {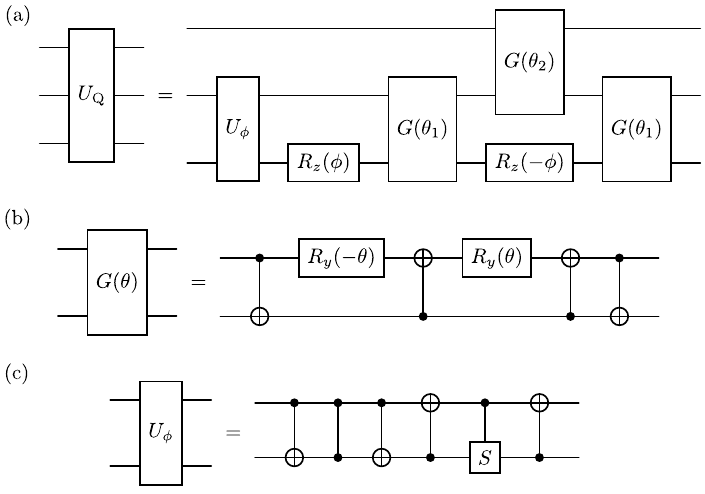}
    \caption{{\bf Encoded qutrit Fourier transform.}
    (a) Decomposition of the qutrit Fourier transform $U_{\mathrm Q}$ in the
    three-qubit unit-Hamming-weight encoding. The circuit consists of
    single-qubit rotations $R_z(\theta)$, three two-qubit Givens rotations
    $G(\theta)$ and the phase gate $U_{\phi}$, obtained using the Reck
    decomposition. The rotation angles are
    $\theta_1=\pi/4$ and
    $\theta_2=\arccos(1/\sqrt{3})$.
    (b) Two-qubit Givens rotation $G(\theta)$ acting on the
    single-excitation subspace.
    (c) Two-qubit phase gate $U_{\phi}$.}
    \label{fig:app-qutrit-fourier}
\end{figure}

The resulting decomposition is shown in
Fig.~\ref{fig:app-qutrit-fourier}. It uses two types of two-qubit operation.
The first is the phase gate
\begin{equation}
    U_{\phi}
    =
    \begin{pmatrix}
        1 & 0 & 0 & 0\\
        0 & i & 0 & 0\\
        0 & 0 & -1 & 0\\
        0 & 0 & 0 & 1
    \end{pmatrix},
    \label{eq:app-uphi}
\end{equation}
written in the computational basis
$\{\ket{00},\ket{01},\ket{10},\ket{11}\}$. The second is the Givens
rotation
\begin{equation}
    G(\theta)
    =
    \begin{pmatrix}
        1 & 0 & 0 & 0\\
        0 & \cos\theta & \sin\theta & 0\\
        0 & -\sin\theta & \cos\theta & 0\\
        0 & 0 & 0 & 1
    \end{pmatrix},
    \label{eq:app-givens}
\end{equation}
which mixes the single-excitation states $\ket{01}$ and $\ket{10}$ while
leaving $\ket{00}$ and $\ket{11}$ unchanged. Consequently, both operations
preserve the encoded qutrit subspace.

Combining these gates with single-qubit $R_z$ rotations implements
$U_{\mathrm Q}$ exactly on the logical qutrit. The decomposition uses three
Givens rotations with
\begin{equation}
    \theta_1=\frac{\pi}{4},
    \qquad
    \theta_2=\arccos\!\left(\frac{1}{\sqrt{3}}\right),
\end{equation}
together with the phase gate $U_{\phi}$ and the indicated single-qubit
rotations. After compilation for the H2 processors, each encoded qutrit
Fourier transform contains 17 single-qubit and 7 two-qubit gates.

\subsection{Initial-state preparation}
\label{sec:Wstate}

The dense reduction of Sec.~\ref{sec:denseAPP} expresses all braiding and
fusion observables required by the protocol as expectation values on the
two-qutrit product state
\begin{equation}
    \ket{\xi}
    =
    \frac{1}{3}
    \sum_{g_1,g_2\in\mathbb{Z}_3}
    \ket{g_1,g_2}.
    \label{eq:app-xi-state}
\end{equation}
Since the state factorises into identical equal superpositions on the two
qutrits, it can be prepared as
\begin{equation}
    \ket{\xi}
    =
    \left(U_{\mathrm Q}\otimes U_{\mathrm Q}\right)
    \ket{e,e},
    \label{eq:app-xi-fourier-preparation}
\end{equation}
where $U_{\mathrm Q}$ is the qutrit Fourier transform defined in
Eq.~\eqref{eq:app-qutrit-qft}.
In the three-qubit encoding
\begin{equation}
    \ket{e}\equiv\ket{100},
    \qquad
    \ket{c}\equiv\ket{010},
    \qquad
    \ket{c^2}\equiv\ket{001},
\end{equation}
the single-qutrit equal superposition is the three-qubit $W$ state,
\begin{equation}
    \ket{W}
    =
    \frac{1}{\sqrt{3}}
    \left(
        \ket{100}
        +
        \ket{010}
        +
        \ket{001}
    \right).
    \label{eq:app-W-state}
\end{equation}
Consequently,
\begin{equation}
    \ket{\xi}
    =
    \ket{W}\otimes\ket{W}.
    \label{eq:app-xi-W-product}
\end{equation}

\begin{figure}
    \centering
    \includegraphics[width=\columnwidth]{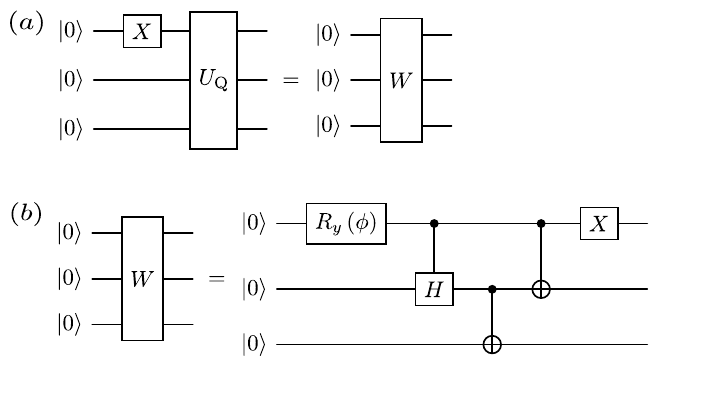}
    \caption{{\bf Preparation of the two-qutrit input state.}
    (a) On the input $\ket{000}$, preparing
    $\ket{e}=\ket{100}$ with an $X$ gate and subsequently applying the
    encoded qutrit Fourier transform $U_{\mathrm Q}$ produces the
    three-qubit $W$ state. This state-specific transformation can be replaced
    by the shorter direct preparation denoted by $W$.
    (b) Circuit decomposition of the direct $W$-state preparation used in
    the experiment. Applying this circuit independently to the two
    three-qubit registers prepares
    $\ket{\xi}=\ket{W}\otimes\ket{W}$.}
    \label{fig:app-W-preparation}
\end{figure}

Rather than implementing the complete encoded Fourier transform solely for
state preparation, we compile a shorter circuit that prepares $\ket{W}$
directly from $\ket{000}$. As shown in
Fig.~\ref{fig:app-W-preparation}, this circuit has the same action on the
specified input as first preparing
$\ket{e}=\ket{100}$ with an $X$ gate and then applying
$U_{\mathrm Q}$. After compilation for the H2 processors, each
three-qubit $W$-state preparation requires four single-qubit and three
two-qubit gates. Preparing the full state $\ket{\xi}$ therefore uses two
identical $W$ circuits acting independently on the two encoded qutrits.

\subsection{Charge-projection operators}
\label{app:projection_operators}

We next describe the ancilla-assisted implementation of the diagonal
projectors $P^{+}$ and $P^{G}$ introduced in
Eqs.~\eqref{eq:Aplus-diagonal} and \eqref{eq:AG-diagonal}. In the
three-qubit unit-Hamming-weight encoding of Eq.~\eqref{eq:encode}, two
logical qutrits are equal if and only if their three corresponding physical
qubits agree pairwise. This condition can therefore be determined using
three parity checks.

\begin{figure}[t!]
    \centering
    \includegraphics[width=\columnwidth]
    {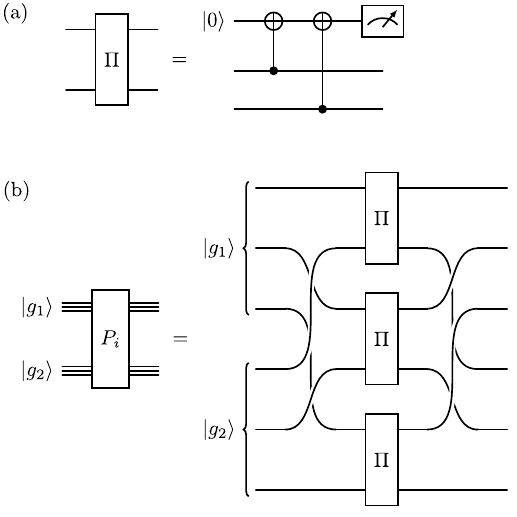}
    \caption{{\bf Ancilla-assisted charge projection.}
    (a) Parity-check circuit $\Pi$. The ancilla records $\ket{0}_{A}$ when
    the two computational qubits have equal occupations and $\ket{1}_{A}$
    when they differ.
    (b) Three parity checks compare the corresponding physical qubits of the
    two encoded qutrits. Within the unit-Hamming-weight code space, equal
    qutrits produce the syndrome $000$, whereas unequal qutrits produce
    $110$, $101$ or $011$. In the coherent implementation, these three
    unequal syndromes are mapped to the common sector flag
    $\ket{11}_{T}$, while the equal syndrome is mapped to
    $\ket{00}_{T}$. The parity ancillas are subsequently uncomputed.
    Post-selection on the two flag outcomes therefore implements the
    $+$- and $G$-sector charge projections without resolving the individual
    unequal-qutrit configurations. The reordering of the circuit lines is
    diagrammatic and does not require physical SWAP gates on the all-to-all
    connected H2 processors.}
    \label{fig:app-projection-operators}
\end{figure}

The elementary parity-check circuit $\Pi$ is shown in
Fig.~\ref{fig:app-projection-operators}(a). For two computational qubits
$\ket{x,y}$ and an ancilla initialised in $\ket{0}_{A}$, it performs
\begin{equation}
    \ket{x,y}\ket{0}_{A}
    \longmapsto
    \ket{x,y}\ket{x\oplus y}_{A}.
    \label{eq:app-parity-check}
\end{equation}
Thus, the ancilla records $\ket{0}_{A}$ when the two qubits have equal
occupations and $\ket{1}_{A}$ when they differ.

Applying $\Pi$ to each pair of corresponding physical qubits produces a
three-bit parity syndrome. Within the logical qutrit code space, equal
qutrits give
\begin{equation}
    000,
\end{equation}
whereas unequal qutrits give one of the three syndromes
\begin{equation}
    110,\qquad 101,\qquad 011.
    \label{eq:app-unequal-syndromes}
\end{equation}
The syndrome $000$ therefore identifies the equal-qutrit subspace and
implements the diagonal projector $P^{+}$. Conjugating this operation by
the qutrit Fourier transforms gives the charge-sector operation
$\mathcal{A}^{+}(v)$, as in Eq.~\eqref{eq:Aplus-diagonal}. The probability
of obtaining the $000$ syndrome is the weight of the input state in the
equal-qutrit subspace. After compilation for the H2 processors, this
implementation uses six computational qubits and three parity ancillas,
together with 66 single-qubit and 34 two-qubit gates.

A direct measurement of the three parity ancillas would distinguish the
syndromes in Eq.~\eqref{eq:app-unequal-syndromes}. This would resolve more
information than the projector $P^{G}=\mathbf{1}_{9}-P^{+}$ and would
therefore destroy coherence between the three components of the unequal
subspace. Such a measurement cannot be used when the charge projection
appears coherently inside a longer ribbon sequence.

To implement $P^{G}$ without revealing which unequal syndrome occurred, we
reversibly compress the parity information into a two-qubit sector register.
A Toffoli-based Boolean circuit maps the syndrome $000$ to the flag
$\ket{00}_{T}$ and all three allowed unequal syndromes to the common flag
$\ket{11}_{T}$. The original parity register is then uncomputed by applying
the inverse parity checks. On logical basis states, the complete coherent
operation therefore acts as
\begin{equation}
\begin{split}
    &\ket{g_1,g_2}
    \ket{000}_{A}
    \ket{00}_{T}
    \\
    &\qquad\longmapsto
    \begin{cases}
        \ket{g_1,g_2}\ket{000}_{A}\ket{00}_{T},
        & g_1=g_2,\\[1mm]
        \ket{g_1,g_2}\ket{000}_{A}\ket{11}_{T},
        & g_1\neq g_2.
    \end{cases}
\end{split}
\label{eq:Acases}
\end{equation}
Here $A$ denotes the three parity ancillas and $T$ the two sector-flag
ancillas. Because the parity register is returned to $\ket{000}_{A}$, the
three unequal syndromes remain coherent and only the distinction between the
$+$ and $G$ sectors is retained.

After conjugation by the Fourier transforms, post-selection on
$\ket{00}_{T}$ implements $\mathcal{A}^{+}(v)$, while post-selection on
$\ket{11}_{T}$ implements $\mathcal{A}^{G}(v)$. The full coherent
construction uses six computational qubits and five ancillas. After
compilation for H2, it contains 84 single-qubit and 52 two-qubit gates.
The rearrangement of qubit lines shown in the circuit is purely diagrammatic:
no SWAP gates are required because the trapped-ion processor provides
all-to-all connectivity.

\subsection{Ribbon operators}
\label{app:ribbon_operators}

The reduced ribbon operators are implemented through the diagonal form given
in Eqs.~\eqref{eq:ribbon-diagonal-1} and
\eqref{eq:ribbon-diagonal-2}. Since the diagonal operator
$\mathcal{D}_{F}$ is non-unitary, the hardware circuit realises the scaled
map $\mathcal{D}_{F}/2$ by embedding it in a unitary dilation with ancillary
qubits.

\begin{figure}
    \centering
    \includegraphics[width=\linewidth]
    {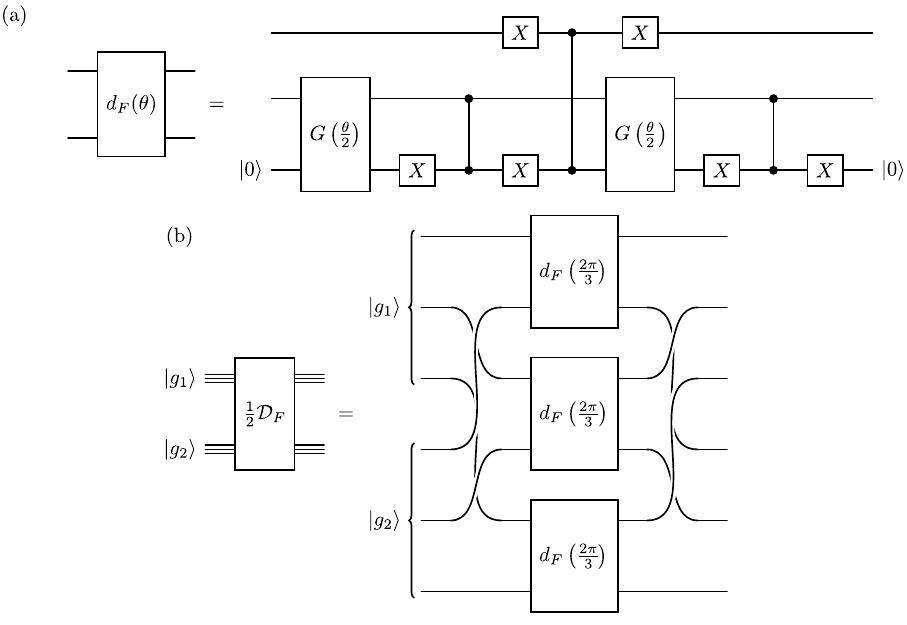}
    \caption{{\bf Ancilla-assisted implementation of the ribbon filter.}
    (a) Unitary dilation of the elementary two-qubit filter
    $d_F(\theta)$. On the branch in which the ancilla returns to
    $\ket{0}_{A}$, the logical state $\ket{01}$ is multiplied by
    $\cos\theta$, while the other computational-basis states are unchanged.
    The gates $G(\theta/2)$ are the Givens rotations defined in
    Eq.~\eqref{eq:app-givens}.
    (b) Implementation of the scaled diagonal ribbon operator
    $\mathcal{D}_{F}/2$. Three filters
    $d_F(2\pi/3)$ act on corresponding pairs of physical qubits in the two
    encoded qutrits. Equal qutrit labels are unchanged, whereas unequal
    labels acquire the factor $-1/2$. Successful operation is associated
    with all three ancillas occupying $\ket{0}_{A}$. The rearrangement of
    the circuit lines is diagrammatic and requires no SWAP gates on the
    all-to-all connected H2 processors.}
    \label{fig:app-ribbon-operators}
\end{figure}

We first introduce the elementary two-qubit filter $d_F(\theta)$ shown in
Fig.~\ref{fig:app-ribbon-operators}(a). For two logical qubits and an
ancilla initialised in $\ket{0}_{A}$, the dilation acts as
\begin{equation}
    U_{d_F}(\theta)
    \ket{\psi}\ket{0}_{A}
    =
    d_F(\theta)\ket{\psi}\ket{0}_{A}
    +
    \ket{\Phi^{\perp}}\ket{1}_{A},
    \label{eq:app-dF-dilation}
\end{equation}
where
\begin{equation}
    d_F(\theta)
    =
    \begin{pmatrix}
        1 & 0 & 0 & 0\\
        0 & \cos\theta & 0 & 0\\
        0 & 0 & 1 & 0\\
        0 & 0 & 0 & 1
    \end{pmatrix}
    \label{eq:app-dF-filter}
\end{equation}
is written in the computational basis
$\{\ket{00},\ket{01},\ket{10},\ket{11}\}$. The successful branch is
identified by the ancilla state $\ket{0}_{A}$, on which only the
$\ket{01}$ component is multiplied by $\cos\theta$.

We choose
\begin{equation}
    \theta=\frac{2\pi}{3},
    \qquad
    \cos\theta=-\frac{1}{2}.
\end{equation}
The diagonal two-qutrit operation is then constructed by applying one such
filter to each pair of corresponding physical qubits in the two encoded
qutrits, as shown in Fig.~\ref{fig:app-ribbon-operators}(b). If the logical
qutrit labels are equal, $g_1=g_2$, the three physical-qubit pairs consist
of one $\ket{11}$ pair and two $\ket{00}$ pairs. No filter therefore
introduces an attenuation, and the state is unchanged. If $g_1\neq g_2$,
the three pairs consist of one $\ket{10}$ pair, one $\ket{01}$ pair and one
$\ket{00}$ pair. Exactly one filter then contributes the factor
$-1/2$. Conditioning on all three operation ancillas occupying
$\ket{0}_{A}$ consequently implements
\begin{equation}
    \frac{1}{2}\mathcal{D}_{F}\ket{g_1,g_2}
    =
    \begin{cases}
        \ket{g_1,g_2}, & g_1=g_2,\\[1mm]
        -\dfrac{1}{2}\ket{g_1,g_2}, & g_1\neq g_2,
    \end{cases}
    \label{eq:app-scaled-DF}
\end{equation}
in agreement with Eq.~\eqref{eq:DF-unscaled}.

Conjugating this diagonal filter by the appropriate qutrit Fourier
transforms gives scaled versions of the two reduced ribbon operators:
\begin{align}
    \frac{1}{2}\mathcal{F}^{G}_{\rho_1}
    &=
    (\mathbf{1}_3\otimes U_{\mathrm Q})
    \frac{\mathcal{D}_{F}}{2}
    (\mathbf{1}_3\otimes U_{\mathrm Q}^{\dagger}),
    \label{eq:app-ribbon-dilation-1}\\
    \frac{1}{2}\mathcal{F}^{G}_{\rho_2}
    &=
    (U_{\mathrm Q}^{\dagger}\otimes\mathbf{1}_3)
    \frac{\mathcal{D}_{F}}{2}
    (U_{\mathrm Q}\otimes\mathbf{1}_3).
    \label{eq:app-ribbon-dilation-2}
\end{align}
The circuits therefore implement $\mathcal{F}^{G}_{\rho_k}/2$, rather than
the unscaled ribbon operator itself. The known factor of two is restored
analytically through the normalization factors used when reconstructing the
braiding and fusion overlaps.

Within the logical two-qutrit subspace, the probability of obtaining the
successful ancilla branch is
\begin{equation}
    p_{\mathrm{succ}}
    =
    P_{\mathrm{eq}}
    +
    \frac{1}{4}P_{\mathrm{neq}},
    \label{eq:app-ribbon-success}
\end{equation}
where $P_{\mathrm{eq}}$ and $P_{\mathrm{neq}}$ are the weights of the input
state in the equal- and unequal-qutrit subspaces. Hence
$1/4\leq p_{\mathrm{succ}}\leq1$. In the fusion circuits, success is selected
by measuring and post-selecting the operation ancillas. In the braiding
Hadamard test, the same ancillas remain part of the coherent unitary
dilation and are not measured.

Each ribbon primitive uses six computational qubits and three operation
ancillas. After compilation for the H2 processors,
$\mathcal{F}^{G}_{\rho_1}/2$ contains 55 single-qubit and 29 two-qubit
gates, whereas $\mathcal{F}^{G}_{\rho_2}/2$ contains 61 single-qubit and
29 two-qubit gates.

\subsection{Hadamard test}
\label{app:hadamard-test}

In Sec.~\ref{sec:braiding-extraction}, the squared braiding phases were
reconstructed from expectation values of the non-unitary two-qutrit maps
$\mathcal{M}_{+}$ and $\mathcal{M}_{\mathbf{1}}$. The map
$\mathcal{M}_{+}$ contains the charge projection
$\mathcal{A}^{+}(v)$, whereas $\mathcal{M}_{\mathbf{1}}$ is obtained by
replacing this projection with the identity. We denote their corresponding
unitary dilations by $M_{+}$ and $M_{\mathbf{1}}$, respectively. The
identity-channel quantity is an auxiliary overlap rather than a physical
anyon braiding phase; it is used to obtain the $G$-sector result through
\begin{equation}
    (R^{G})^{2}
    =
    (R^{\mathbf{1}})^{2}
    -
    (R^{+})^{2},
    \label{eq:app-RG-subtraction}
\end{equation}
as in Eq.~\eqref{eq:RG-measurement}. Here we describe the state-specific
controlled circuits used to perform the Hadamard tests for these dilations.

\begin{figure*}[!ht]
    \centering
    \includegraphics[width=\linewidth]{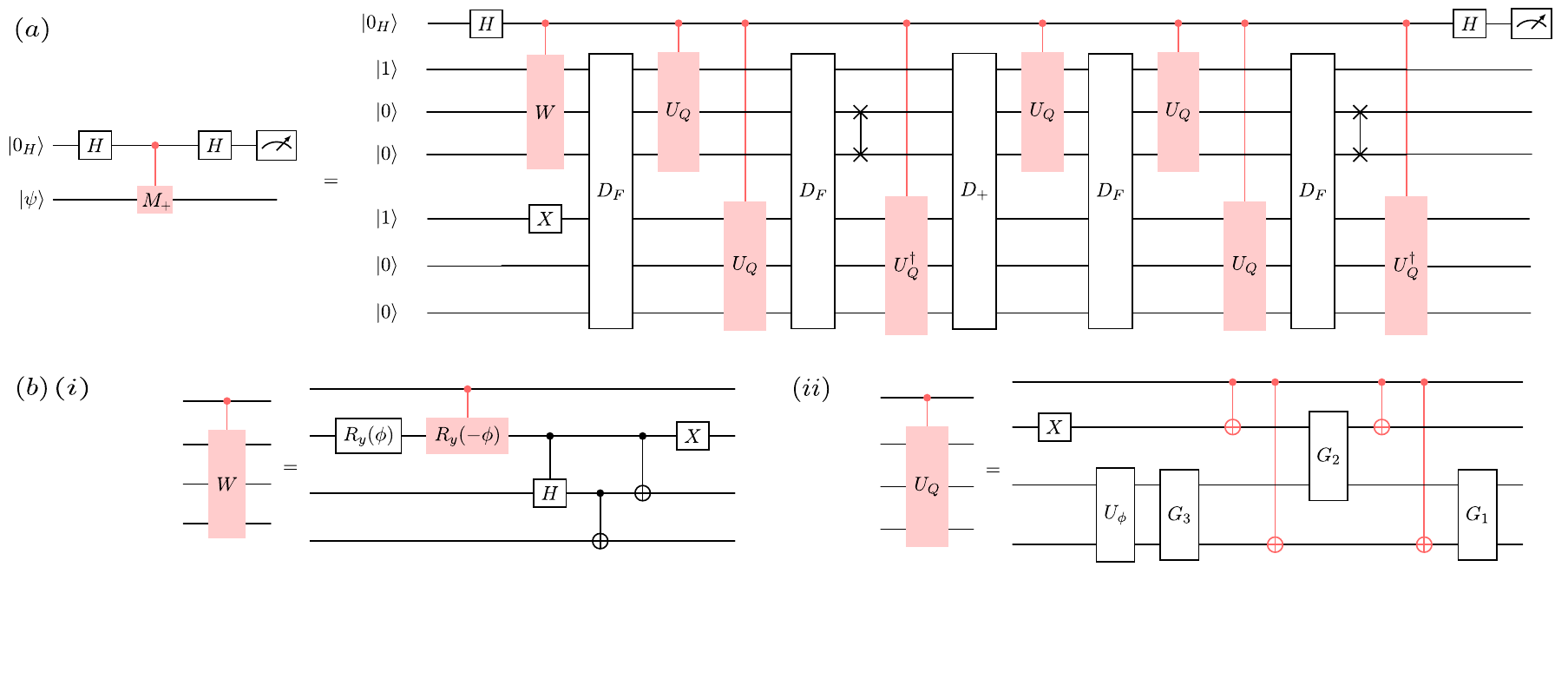}
    \caption{{\bf State-specific controlled Hadamard test.}
    (a) Left: standard Hadamard test for extracting
    $\operatorname{Re}[\bra{\Psi}M\ket{\Psi}]$ from the Hadamard-ancilla
    probabilities. Right: optimized implementation of the $+$-sector
    dilation $M_{+}$ for the fixed input
    $\ket{\Psi_{+}}=\ket{100,100}\ket{0}^{\otimes15}_{A}$.
    Gates shown in red are conditioned on the Hadamard ancilla, whereas the
    diagonal ribbon blocks $D_F$, the charge block $D_{+}$ and the indicated
    permutations are applied unconditionally. The two circuits produce the
    same Hadamard-ancilla statistics for the specified input, but need not
    define the same controlled unitary on arbitrary states. Ancillas internal
    to the $D_F$ and $D_{+}$ blocks are not shown.
    (b) Decompositions of the controlled operations used in panel (a):
    (i) controlled-$W$ state preparation and
    (ii) controlled qutrit Fourier transform $U_{\mathrm Q}$, implemented
    using the phase gate $U_{\phi}$ and Givens rotations
    $G_1$, $G_2$ and $G_3$.}
    \label{fig:Htest}
\end{figure*}

For a unitary operator $M$ and an input state $\ket{\Psi}$, the standard
Hadamard test is shown on the left-hand side of
Fig.~\ref{fig:Htest}(a). After preparing the Hadamard ancilla in
$(\ket{0}_{H}+\ket{1}_{H})/\sqrt{2}$ and applying $M$ conditional on
$\ket{1}_{H}$, the joint state is
\begin{equation}
    \frac{1}{\sqrt{2}}
    \left(
        \ket{0}_{H}\ket{\Psi}
        +
        \ket{1}_{H}M\ket{\Psi}
    \right).
    \label{eq:app-hadamard-intermediate}
\end{equation}
Applying a final Hadamard gate and measuring the ancilla gives
\begin{equation}
    p^{\mathrm{Re}}_{H}(0)-p^{\mathrm{Re}}_{H}(1)
    =
    \operatorname{Re}
    \left(
        \bra{\Psi}M\ket{\Psi}
    \right).
    \label{eq:standard-hadamard-real}
\end{equation}
Inserting an $S^{\dagger}$ gate immediately before the final Hadamard gives
the imaginary quadrature,
\begin{equation}
    p^{\mathrm{Im}}_{H}(0)-p^{\mathrm{Im}}_{H}(1)
    =
    \operatorname{Im}
    \left(
        \bra{\Psi}M\ket{\Psi}
    \right),
    \label{eq:standard-hadamard-imaginary}
\end{equation}
with the phase convention used in Eq.~\eqref{eq:Rplus-measurement}.

The unitary dilations contain known scale factors arising from the
implementation of the non-unitary ribbon and projection operators.
Accounting for these factors gives
\begin{equation}
\begin{split}
    \operatorname{Re}\!\left[(R^{+})^{2}\right]
    &=
    8\left[
        p^{\mathrm{Re}}_{H}(0)
        -
        p^{\mathrm{Re}}_{H}(1)
    \right],
    \\
    \operatorname{Im}\!\left[(R^{+})^{2}\right]
    &=
    8\left[
        p^{\mathrm{Im}}_{H}(0)
        -
        p^{\mathrm{Im}}_{H}(1)
    \right],
\end{split}
\label{eq:app-hadamard-Rplus}
\end{equation}
equivalent to Eq.~\eqref{eq:Rplus-measurement} because
$p_H(0)+p_H(1)=1$. The same normalization convention is used for the
identity-channel overlap $(R^{\mathbf{1}})^2$.

For the $+$-sector circuit, the input to the dilation is
\begin{equation}
    \ket{\Psi_{+}}
    =
    \ket{100,100}
    \ket{0}^{\otimes15}_{A},
    \label{eq:Psi-plus-hadamard}
\end{equation}
where the first six qubits encode the two qutrits and the remaining
15 qubits are operation ancillas. A direct implementation of
controlled-$M_{+}$ would require promoting every constituent gate of the
dilation to a controlled operation. Since $M_{+}$ contains several qutrit
Fourier transforms, ribbon filters and charge-projection blocks, this would
produce a substantial two-qubit-gate overhead.

A general controlled implementation is unnecessary because the Hadamard test
is applied only to the fixed input state in
Eq.~\eqref{eq:Psi-plus-hadamard}. We instead construct a state-specific
controlled circuit $\widetilde{M}_{+}$ satisfying
\begin{equation}
\begin{split}
    \widetilde{M}_{+}
    \frac{
        \ket{0}_{H}\ket{\Psi_{+}}
        +
        \ket{1}_{H}\ket{\Psi_{+}}
    }{\sqrt{2}}
    =
    \frac{
        \ket{0}_{H}\ket{\Psi_{+}}
        +
        \ket{1}_{H}M_{+}\ket{\Psi_{+}}
    }{\sqrt{2}}.
\end{split}
\label{eq:state-specific-control}
\end{equation}
Thus, the $\ket{1}_{H}$ branch undergoes the complete dilation $M_{+}$,
whereas the $\ket{0}_{H}$ branch returns to the original state
$\ket{\Psi_{+}}$. Equation~\eqref{eq:state-specific-control} establishes
equivalence only for this specified input; $\widetilde{M}_{+}$ need not
coincide with a general controlled-$M_{+}$ operation on arbitrary states.

The optimized construction is shown on the right-hand side of
Fig.~\ref{fig:Htest}(a). Operations that already act as the identity on the input state $\ket{100}\ket{100}$
are left uncontrolled, and only the state-preparation and
basis-change gates needed to distinguish the two Hadamard branches are conditioned on
the Hadamard ancilla. In particular, the diagonal ribbon blocks $D_F$, the
charge block $D_{+}$ and the indicated permutations are applied without an
additional Hadamard control. Controlled operations are shown in red in the
figure. In the decomposition used here, the dilation $M_{+}$ contains
311 gates, but only 29 operations need to be conditioned on the Hadamard
ancilla.

The controlled state-preparation and Fourier-transform circuits are given in
Fig.~\ref{fig:Htest}(b). Panel (b)(i) shows the controlled-$W$
state-preparation circuit, while panel (b)(ii) shows the controlled qutrit
Fourier transform $U_{\mathrm Q}$. The latter is expressed using the phase
gate $U_{\phi}$ and the three instances $G_1$, $G_2$ and $G_3$ of the
Givens rotation defined in
Eqs.~\eqref{eq:app-uphi} and \eqref{eq:app-givens}.

The same state-specific strategy is used for the identity-channel dilation
$M_{\mathbf{1}}$. In this case the $D_{+}$ block is omitted and the input is
\begin{equation}
    \ket{\Psi_{\mathbf{1}}}
    =
    \ket{100,100}
    \ket{0}^{\otimes12}_{A}.
    \label{eq:Psi-identity-hadamard}
\end{equation}

The operation ancillas form part of the coherent unitary dilations and are
neither measured nor post-selected in the Hadamard-test experiments. Only
the Hadamard ancilla is used in the reconstruction of the real and imaginary
quadratures. Consequently, every experimental shot contributes to the
measured expectation value.

\putbib
    
\end{bibunit}

\end{document}